\documentclass[structabstract]{aa}  
\usepackage{graphicx}
\usepackage{txfonts}
\usepackage{natbib}
\bibpunct{(}{)}{;}{a}{}{,} 
\newcommand{\Hii}{{\sc H$\,$ii} }

\begin{document}
   \title{The escape of ionising radiation from high-redshift dwarf galaxies}

   \author{J.-P. Paardekooper
          \inst{1,2}
          \and
          F.I. Pelupessy
          \inst{2}
          \and
          G. Altay
          \inst{3}
          \and
          C.J.H. Kruip
          \inst{2}
          }

   \institute{
         Max-Planck-Institut f\"{u}r extraterrestrische Physik, Giessenbachstra\ss e, 85748 Garching, Germany\\
         \email{jppaarde@mpe.mpg.de}
         \and
         Leiden Observatory, Leiden University, Postbus 9513, 2300RA Leiden, The Netherlands
         \and
         Institute of Computational Cosmology, Department of Physics, University of Durham, Science Laboratories, SouthRoad, Durham DH1 3LE
            }

   \date{Received 7 March 2011 / Accepted 14 April 2011}

 
  \abstract
   {The UV escape fraction from high-redshift galaxies plays a key role in models of cosmic reionisation. Because it is currently not possible to deduce the escape fractions during the epoch of reionisation from observations, we have to rely on numerical simulations.}
   {We aim to better constrain the escape fraction from high-redshift dwarf galaxies, as these are the most likely sources responsible for reionising the Universe.}
   {We employ a N-body/SPH method that includes realistic prescriptions for the physical processes that are important for the evolution of dwarf galaxies. These models are post-processed with radiative transfer to determine the escape fraction of ionising radiation. We perform a parameter study to assess the influence of the spin parameter, gas fraction and formation redshift of the galaxy and study the importance of numerical parameters as resolution, source distribution and local gas clearing.}
   {We find that the UV escape fraction from high-redshift dwarf galaxies that have formed a rotationally supported disc lie between $10^{-5}$ and 0.1. The mass and angular momentum of the galaxy are the most important parameters that determine the escape fraction. We compare our results to previous work and discuss the uncertainties of our models.}
   {The low escape fraction we find for high-redshift dwarf galaxies is balanced by their high stellar content, resulting in an efficiency parameter for stars that is only marginally lower than the values found by semi-analytic models of reionisation. We therefore conclude that dwarf galaxies play an important role in cosmic reionisation also after the initial starburst phase, when the gas has settled into a disc.}


   \maketitle
%

\section{Introduction}

Evidence from WMAP data \citep{Spergel:2007p2960,Page:2007p2959,Komatsu:2010p2993} and observations of the Lyman alpha forest from high-redshift quasars (\citet{Becker:2001p2962,Fan:2001p2963,Fan:2006p2964}, for a review see \citet{Fan:2006p2961}) show that the Universe was likely reionised between redshifts 11 and 6. Extrapolation of the observed quasar and stellar luminosity functions to high redshift indicate that the majority of photons contributing to reionisation originated in galaxies as opposed to quasars \citep{Madau:1999p2786,Fan:2002p2792,Yan:2004p2966,Srbinovsky:2007p2801,Bouwens:2010p3250}. However, the nature of these stellar sources and the mass range of galaxies that host them remains uncertain. 

In the standard cold dark matter paradigm, most ionising photons produced during the epoch of reionisation originate in  galaxies with masses between $10^{8} - 10^{10} M_{\odot}$ (e.g. \citet{Barkana:2001p19}). This picture is corroborated by semi-analytic models of reionisation showing that if the Universe was reionised by stellar sources, a significant population of low mass galaxies must exist to drive reionisation \citep{Choudhury:2008p2944}. Recent observations of the luminosity function at $z \sim 7$ also suggest a large population of low mass galaxies is necessary in order to produce enough photons to reionise the Universe
(e.g.\citet{Oesch:2009p3239,Richard:2008p3272,Bouwens:2010p3250}). However, no direct evidence of these galaxies' contributions to reionisation exists. The number of photons contributed depends on the intrinsic number of ionising photons produced in the galaxies, and the fraction able to escape from the galaxies. The star formation rate and initial mass function, both of which are uncertain at high redshift, determine the first of these factors. The second factor, the escape fraction $f_{\mathrm{esc}}$, also has a large uncertainty in reionisation modelling. Understanding this parameter in isolated galaxy models is the focus of this work.  

The photo-ionisation rate at redshifts 5-6 inferred from the Ly-$\alpha$ opacity of the IGM indicates that the observed population of galaxies and quasars is capable of maintaining the IGM in its ionised state if the escape fraction is higher than 20\% \citep{Bolton:2007p2789}. Although their method to determine the photo-ionisation rate only works for an IGM that is already ionised, i.e., for redshifts below 6, extrapolation to higher redshifts suggests that the reionisation epoch was extended and occurred in a photon-starved manner. Unless the ionising emissivity per unit comoving volume is substantially higher during reionisation than it is now, 1.5 - 3 photons per hydrogen atom are produced over the lifetime of the Universe. Recent observations of galaxies at redshifts within the epoch of reionisation show that the observed population of galaxies can only drive reionisation if the escape fraction is higher than 20\% \citep{Labbe:2010p2967}, or even as high as 60\% \citep{Bouwens:2010p3250}. The abundance of unobserved low luminosity sources depends heavily on the parameterisation of the faint-end slope of the luminosity function.  However, it is interesting to ask whether escape fractions of 20-60 \% in high-redshift galaxies are realistic. For this reason, a lot of effort has gone into observationally determining how many UV photons are able to escape from galaxies at high redshift.

Determining the escape fraction observationally is a difficult task even in the local Universe. For the Milky Way, the escape fraction has been estimated by modelling observed emission measures and kinematics of the Magellanic Stream assuming photoionisation due to hot, young stars in the galactic disc.  The result is escape fractions $ \sim 6\%$ perpendicular to the galactic disc and $ \sim 1-2\%$ when averaged over solid angle \citep{BlandHawthorn:1999p2971,BlandHawthorn:2001p2989}. Determining the escape fraction from galaxies other than the Milky Way is complicated by the fact that the intrinsic number of ionising photons that is produced in the galaxy is unknown. An often used method is to measure the flux at a specific wavelength (usually 900 \AA) and then use models to determine the total number of Lyman continuum photons from that flux \citep{Leitherer:1995p2983}. This requires knowledge of the star formation history and IMF in the galaxy. At low redshift one can use the H$\alpha$ flux to constrain the total number of Lyman continuum photons produced in the galaxy. At high redshift, this becomes increasingly difficult as the Balmer lines shift to the near-IR. Therefore, the flux at 1500 {\AA } is often used instead. However, the 1500 {\AA } line is subject to significant dust
attenuation, so it is difficult to determine the intrinsic Lyman continuum luminosity. For that reason, \citet{Steidel:2001p2973} defined the relative escape fraction that is independent of the dust attenuation:
\begin{equation} 
f_{\mathrm{esc,rel}} = \frac{L_{1500}/L_{900}}{f_{1500}/f_{900}} \mathrm{exp}( \tau_{\mathrm{IGM},900}).
\end{equation}
Here, $f_{1500}$ is the observed flux at 1500 {\AA } and $f_{900}$ the observed flux at 900 \AA. The line-of-sight opacity of the IGM to Lyman continuum photons $\tau_{\mathrm{IGM},900}$ can be estimated empirically or through simulations, while the ratio between intrinsic luminosity above and below the Lyman break $L_{1500}/L_{900}$ needs to be estimated from stellar population synthesis models. The relation between the absolute escape fraction and relative escape fraction is then
\begin{equation} 
f_{\mathrm{esc}} = 10^{-0.4 A(1500)} f_{\mathrm{esc,rel}},
\end{equation}
where $A(1500) = 10.33E(B-V)$ is the dust extinction at 1500 \AA. Since in this work we are primarily interested in the absolute escape fraction, when necessary we convert values for the relative escape fraction in the quoted literature to absolute escape fractions using the median value $E(B-V) = 0.15$ \citep{Shapley:2006p2720,Siana:2007p2724}.

Attempts to observe the Lyman continuum flux of galaxies in the local Universe yielded null results, thus placing only upper limits on the escape fraction. The escape fraction from five local starburst galaxies that were selected for a high predicted escape fraction were found to be less than 3 - 10\% \citep{Hurwitz:1997p2979,Deharveng:2001p2980}. For the local extreme starburst dwarf galaxy Haro 11, that was selected to be representative for dwarf starbursts at high redshift, \citet{Bergvall:2006p2982} found the escape to be less than 10\%, but reanalysis by \citet{Grimes:2007p2984} showed that for the same galaxy $f_{\mathrm{esc}} < 2\%$. \citet{Heckman:2001p2981} estimated the local {\sc H i} column density in 5 local starbursts from {\sc C ii} and {\sc O i} absorption lines and found escape fractions less than 6\% in these galaxies. Null results for Lyman continuum flux searches have also been reported at redshifts around 1. Using imaging techniques \citet{Malkan:2003p2985} placed upper limits of 6\% on the escape fractions for 11 bright blue galaxies at $z \sim  1$. \citet{Siana:2007p2724} added 21 galaxies at similar redshifts to this sample, for which no escaping Lyman continuum photons were found. Stacking the images results in an upper limit to the escape fraction of $ \sim 2$\%. Deep far-UV images of 15 galaxies at $z \sim  1$ obtained by \citet{Siana:2010p2828} have also yielded null detections, constraining the escape fraction to be less than 1\% if the majority of galaxies have non-zero escape fractions. Alternatively, if no radiation escapes from the majority of galaxies, no more than 8\% of the galaxies can have escape fractions higher than 12\%. 

Searches for Lyman continuum emission at higher redshifts have been more successful. \citet{Steidel:2001p2973} found Lyman continuum emission in a stack of 29 galaxies at redshift $z \sim 3.4$, resulting in escape fractions of $ \sim 10$\%. However,  \citet{Giallongo:2002p2722} found no significant Lyman continuum emission from the two brightest galaxies in the same sample, placing an upper limit of $ \sim 4$\% on the escape fraction. For a sample of 27 spectroscopically identified galaxies at redshifts between 1.9 and 3.5, \citet{FernandezSoto:2003p2723} found a similar upper limit. \citet{Shapley:2006p2720} were the first to detect Lyman continuum emission from individual galaxies in a sample of 14 $z \sim 3$ star-forming galaxies. In two out of the 14 galaxies, the majority of produced Lyman continuum photons escape from the galaxy, while the sample average shows an escape fraction of $ \sim 4$\%. \citet{Iwata:2009p2976} find escape fractions ranging from 4\% to 20\% for galaxies in a protocluster at $z \sim 3$, an environment that may not be representative for galaxies at that epoch. An entirely different way of measuring escape fractions that does not suffer from the systematic uncertainties of the methods described earlier is using the afterglow of gamma ray bursts. In a large sample of gamma ray bursts at high redshift \citet{Chen:2007p2974} and \citet{Fynbo:2009p2975} measured escape fractions of less than 7\% . However, the sample of gamma ray burst hosts might be biased due to lack of S/N in optically thin sight lines, resulting in a bias towards a low value of the escape fraction.

The lack of Lyman continuum detections at low redshift and the positive detections at higher redshift might be an indication that escape fractions are higher at high redshift.  However, the samples are very small so it could very well be that the low-redshift counterparts of the high redshift galaxies have not been targeted yet. The high redshift surveys are most likely biased towards the bright end of the luminosity function and the observed objects may be untypical extreme objects. A generic conclusion on the escape fractions at high redshift can therefore not be drawn from these data sets. Larger samples are needed to resolve this issue. Observations might also give a clue on how the photons escape from the galaxy. The result that 2 out of 14 galaxies at $z \sim 3$ have very high escape fractions could indicate that the escape of ionising radiation is highly inhomogeneous and is only detected when the photons are emitted in our line of sight. Observations of the star formation surface density in galaxies that show high escape fractions could provide information on the  part of the galaxy from which radiation is escaping. This would provide insight into the physical processes at work when ionising radiation escapes.

Early attempts of theoretical estimates of the escape fraction relied on simplifying assumptions for the gas distribution in a galaxy. \citet{Dove:2000p2725} studied the formation of superbubbles in a uniform medium and found that around 7\% of the ionising radiation produced by OB associations could escape from a galaxy like the Milky Way. For higher redshifts, \citet{Wood:2000p2728} found that almost no ionising radiation escapes from disc galaxies at redshift 10. \citet{Ricotti:2000p2729} showed that for spheroidal galaxies the maximum escape fraction is 10\%, dropping steeply with increasing mass and redshift. However, \citet{Ciardi:2002p2726} showed that escape fractions are highly dependent on the density distribution in the
galaxy and \citet{Clarke:2002p2727} argued that porosity of the interstellar medium (ISM) caused by supernovae has a profound impact on the escape fraction by providing channels through which radiation can escape. This was investigated by \citet{Fujita:2003p2711} who showed that shells blown by supernovae in starburst dwarf galaxies create chimneys through which radiation escapes, resulting in escape fractions up to 20\%. All these models have in common that the gas distribution in the galaxy is postulated and does not evolve under the influence of the physical processes in the galaxy. Although this provides valuable insight into the dependence of the escape fraction the gas distribution, a realistic estimate requires that the interplay between stars and gas in the galaxy be taken into account. 

In recent years simulations of large scale structure and galaxy formation have been conducted, providing realistic initial conditions for higher resolution re-simulations of select galaxies to determine the UV escape fraction. \citet{Razoumov:2006p2710,Razoumov:2007p2709,Razoumov:2010p1840} extracted galaxies from cosmological simulations and calculated escape fractions using a ray-tracing radiative transfer method on the SPH particles. They found a strong redshift dependence of the escape fraction, ranging from 1-2\% at $z \sim 2$ and 8 - 10\% at $z \sim 3$ up to around 80\% at $z \sim 10$. Using AMR radiation hydrodynamics simulations of primordial starburst dwarf galaxies formed at redshift 8, \citet{Wise:2009p1755} found escape fractions around unity. On the other hand, \citet{Gnedin:2008p2449} found significantly lower escape fractions of 1-3\% for galaxies at redshift $z = 3 - 9$, with almost no redshift dependence with a moment
radiative transfer method coupled to AMR hydrodynamics. Contrary to the above results, they found escape fractions decline sharply in low mass galaxies. \citet{Yajima:2009p2719} post-processed an SPH simulation of an isolated, supernova-dominated galaxy at $z = 3.7 - 7$ with a ray-tracing radiative transfer algorithm to find escape fractions of 20 - 60\% that were highly sensitive to dust extinction. Using a larger sample of galaxies \citet{Yajima:2011p2925} found a high dependence of the escape fraction on galaxy mass, ranging from up to 70\% for low mass galaxies to around 7\% for high mass galaxies. Contrary to the other studies, they find no variation of the escape fraction with redshift between $z = 6$ and $z = 3$. An entirely different approach was taken by \citet{Wyithe:2010p2422}, who used star formation rates derived from gamma ray burst afterglows to break the degeneracy between star formation rate and escape fraction in their semi-analytic models of reionisation. They showed that for plausible reionisation scenarios the escape fraction must be around 5\% at $z > 6$, significantly lower than what is found in the numerical simulations.

These numerical simulations have shown that the escape fraction is highly dependent on the physical processes that shape the galaxy. The intricate interplay between star formation and feedback and the gas inside the galaxy ultimately determines how many photons can escape. However, realistic modelling of galaxies is computationally challenging and requires accurate treatment of many
physical processes in combination with high resolution to resolve all relevant scales. In this work, we apply a numerical method that includes a realistic treatment of the physics inside galaxies and has been shown to reproduce properties like morphology, star formation rate and the spatial pattern of star formation in local dwarf irregular galaxies very well \citep{Pelupessy:2004p2998}. Our primary interest lies in radiation escaping from the main sources driving cosmic reionisation, high-redshift dwarf galaxies. To get an unbiased picture of how the physics in the galaxy itself influences the escape fraction, we only study galaxies in isolation in this work. This also allows us to easily use a fixed resolution for different galaxy masses, something that cannot be done with galaxies extracted from large scale structure simulations. A drawback of this approach is that we neglect infalling gas from outside the halo and merger events, which can increase the star formation rate and production of ionising photons as well as change the gas distribution. These issues will be addressed in future work. In this work, we will explore trends in the escape fraction by varying several relevant physical parameters, such as the halo mass, spin parameter, gas fraction and redshift.

This paper is organised as follows. In Sect.~\ref{section_method} we describe the numerical methods used to simulate the high-redshift dwarf galaxies and the ionising radiation. In Sect.~\ref{section_results} we present the results of the simulations, followed by the implications for reionisation models and a comparison to previous work in Sects.~\ref{section_implications} and \ref{section_discussion}. We end with conclusions in Sect.~\ref{section_conclusions}.


\section{Method}\label{section_method}

The method used for modelling the dwarf galaxies is described in detail in \citet{Pelupessy:2004p2998} and \citet{Pelupessy:2005p3001}. In short, we use an SPH code that follows the evolution of star particles and gas particles in a static dark matter potential. The two-phase nature of the ISM is reproduced in our models as a natural result of the physics included. Star formation follows a Jeans criterion. These models are post-processed with the {\sc SimpleX} radiative transfer method \citep{Paardekooper:2010p2772} to calculate the fraction of ionising radiation that escapes the galaxies.


\subsection{Initial conditions}

In this work, we study the escape fraction from single, isolated dwarf galaxies that have formed a rotationally supported disc. We thereby neglect external gas inflow and feedback effects from nearby galaxies. Although this is clearly an approximation, it gives us the opportunity to isolate and test the influence of simple physical parameters on the escape fraction. It is currently unclear whether high-redshift dwarf galaxies are indeed able to form discs. On the one hand, conservation of angular momentum will lead to disc formation, while on the other hand feedback effects from supernova (SN) explosions may prevent the gas from settling into a disc. The main uncertainty in this issue is the strength of supernova feedback at high redshift, which is currently under debate (e.g. \citet{Agertz:2010p3154}). We will show that SN feedback also has a large effect on the escape fraction (Sect.~\ref{section_results_escape}). In this work we assume that the gas and stars are able to form a disc-like structure and begin our simulations at that point.  Note however that, depending on the physical parameters of the galaxy model, feedback effects can result in the disc evolving into irregular galaxy morphologies. We plan to study the escape fraction during the initial assembly phase (prior to the formation of the disc) of these galaxies in future work.

Each galaxy starts out as a disc of gas and stars that resides in a dark matter halo, constructed using the analytic disk galaxy models of \citet{Mo:1998p2918}. The initial conditions are generated in a similar fashion as described in \citet{Springel:2005p2919}. The dark matter halo follows a profile proposed by \citet{Hernquist:1990p3065}:
\begin{equation} 
\rho(r) = \frac{M_{\mathrm{halo}}}{2 \pi} \frac{a}{r(r+a)^{3}},
\end{equation}
where $M_{\mathrm{halo}}$ is the total mass of the dark matter halo and $a$ is the scale length. This dark matter profile can be related to the widely used NFW profile \citep{Navarro:1997p3008} by expressing $a$ in terms of the scale length $r_{\mathrm{s}}$ of the NFW profile:
\begin{equation} 
a = r_{\mathrm{s}} \sqrt{2 \left( \ln(1+c) - c/(1+c) \right)}.
\end{equation}
Here, $c$ is the concentration factor, defined as $c = r_{200}/r_{\mathrm{s}}$ with $r_{200}$ the radius at which the mean enclosed dark matter density is 200 times the critical density. By defining $a$ this way the dark matter profile agrees with the NFW profile in the inner parts of the halo, while in the outer parts it declines more quickly, so that the total mass converges. This allows us to create isolated haloes without the need to truncate the density distribution at large radii. We represent the dark halo by a static potential. This is deemed sufficient for the dynamical modelling presented here, as we evolve the galaxy in isolation and the perturbations in the gaseous and stellar disk are expected to have only minor impact on the halo structure.

All the baryons are initially in a disc. The total mass of the disc is taken to be a fraction of the total mass of the galaxy, $M_{\mathrm{d}} = f_{\mathrm{b}}M_{\mathrm{tot}}$, with $M_{\mathrm{tot}} = M_{\mathrm{d}} + M_{\mathrm{halo}}$. We take the baryon fraction $f_{\mathrm{b}}= 0.041$ in all galaxy models. The disc consists of stars and gas, so $M_{\mathrm{d}} = M_{\mathrm{gas}} + M_{\star}$, where $M_{\mathrm{gas}} = f_{\mathrm{gas}} M_{\mathrm{d}}$. The fraction of disc mass that is in gas $f_{\mathrm{gas}}$ is a free parameter in the models, and can be taken as a proxy for how far the system has evolved. The surface density of the stars is modelled by an exponential profile of scale length $h_{\star}$:
\begin{equation}\label{eq_surface_density_stars} 
\Sigma_{\star}(R) = \frac{M_{\star}}{2 \pi h_{\star}^{2}}\exp{(-R/h_{\star})},
\end{equation}
while the surface density of the gas disc has a more extended profile:
\begin{equation}\label{eq_surface_density_gas} 
\Sigma_{\mathrm{gas}}(R) = \frac{\Sigma_{\mathrm{gas},0}}{1 + R/h_g},
\end{equation}
where $h_g$ is the scale length of the gas disc. If we assume that the disc is centrifugally supported and that the angular momentum of the material that forms the disc is conserved there is a direct relation between the scale lengths of the gas and stellar discs and the spin parameter of the galaxy $\lambda$. The latter is defined in its usual sense:
\begin{equation} 
 \lambda = \frac{J \vert E \vert ^{1/2}}{G M^{5/2}},
\end{equation}
where $J$ is the angular momentum and $E$ is the total energy of the halo. The spin parameter is a free parameter of our galaxy models, which then sets the scale length of the galactic disc. 

The three-dimensional stellar density follows the profile of an isothermal sheet with radially constant scale length $h_{\mathrm{z}}$ 
\begin{equation} \label{eq_stellar_density}
\rho_{\star}(R,z) = \frac{\Sigma_{\star}}{2h_{\mathrm{z}}} \exp{(-R/h_{\mathrm{s}})} \, \, \mathrm{sech}^{2}(z/h_{\mathrm{z}}).
\end{equation}
It is not possible to prescribe a vertical profile for the gas disc in the same manner, because for the gas disc the vertical profile is governed by a combination of self-gravity and pressure of the gas. In our models we assume that the gas disc is in hydrostatic equilibrium, so that we can write
\begin{equation} 
\frac{\partial \rho_{\mathrm{gas}}}{\partial z} = -\frac{\rho_{\mathrm{gas}}^{2}}{\gamma P} \frac{\partial \Phi}{\partial z},
\end{equation}
where $\gamma$ is the local polytropic index of the equation of state and $\Phi$ is the total gravitational potential. The solution of this equation is determined by the integral constraint
\begin{equation} 
\Sigma_{\mathrm{gas}}(R) = \int \rho_{\mathrm{gas}}(R,z) \mathrm{d}z,
\end{equation}
with $\Sigma_{\mathrm{gas}}(R)$ given by Eq. (\ref{eq_surface_density_gas}). For more details on how the potential and the resulting gas distribution are calculated self-consistently we refer the reader to \citet{Springel:2005p2919}.

We have run different galaxy models varying the spin parameter, total mass of the galaxy, fraction of disc mass in gas, and the formation redshift. The last determines the extent of the dark matter halo and thus the surface density for a given mass. The metallicity is kept constant in all models at a value of 0.2 solar. The galaxy models and their parameters are listed in Table \ref{table_models}.
\begin{table} 
\caption{Galaxy model parameters}             
\label{table_models}      
\centering                          
\begin{tabular}{l l l l l l}        
\hline\hline                 
Galaxy & $M (M_{\odot})$ & $N_{\mathrm{part}}$ & $\lambda$ & $z_{\mathrm{coll}}$ & $f_{\mathrm{gas}}$ \\    
\hline                        
   1 & $10^{8}$ & 20000 & 0.05 & 11 & 0.5 \\      
   2 & $10^{8}$ & 20000 & 0.025 & 11 & 0.5 \\
   3 & $10^{8}$ & 20000 & 0.1 & 11 & 0.5 \\
   4 & $10^{8}$ & 20000 & 0.05 & 11 & 0.2 \\     
   5 & $10^{8}$ & 20000 & 0.05 & 11 & 0.8 \\     
   6 & $10^{8}$ & 20000 & 0.05 & 9 & 0.5 \\      
   7 & $10^{8}$ & 20000 & 0.05 & 7 & 0.5 \\ 
   8 & $10^{9}$ & 200000 & 0.05 & 8  & 0.5 \\
   9 & $10^{9}$ & 200000 & 0.025 & 8 & 0.5 \\ 
   10 & $10^{9}$ & 200000 & 0.1 & 8 & 0.5 \\ 
   11 & $10^{9}$ & 200000 & 0.05 & 8  & 0.2\\
   12 & $10^{9}$ & 200000 & 0.05 & 8  & 0.8\\
   13 & $10^{9}$ & 200000 & 0.05 & 9  & 0.5 \\
   14 & $10^{9}$ & 200000 & 0.05 & 7  & 0.5 \\
\hline                                   
\end{tabular}
\end{table}
All models were evolved for 550 Myr, which means the simulations end around redshift 6. It is highly unlikely that over the course of this time our assumption that the galaxies are in isolation remains valid. However, in this work we are mainly interested in the mean escape fraction, and this simulation time guarantees enough statistics for reliable time averaging.
 

\subsection{The ISM model}\label{section_ism}

Star formation in galaxies is governed by the complex interplay between the interstellar medium and stars. The ISM provides the material from which stars form, while stars influence the ISM with their UV-radiation, stellar winds, and supernova explosions. It is therefore crucial for realistic models of galaxies to include a model for the ISM that captures its observed properties, in order to treat the star formation inside the galaxy correctly. 

The diffuse gas in the ISM is observed to be in three dominant phases: a cold phase ($T \sim 100$ K), a warm phase ($T \sim 10^4$ K) and a hot phase ($T \sim 10^6$ K) \citep{McKee:1977p3236}. Gas in the cold phase is neutral, while gas in the hot phase is ionised. The warm phase consists of both neutral and ionised gas. Star formation is mainly determined by gas in the warm and the cold phase, so for a realistic treatment of star formation it is important to model these two phases correctly.

Our model for the ISM includes the essential physics for the cold and warm neutral medium \citep{Pelupessy:2005p3001} and is qualitatively similar to that described in \citet{Wolfire:1995p2765}. The two-phase nature of the ISM is maintained by a combination of metal cooling, cosmic ray ionisation, and photo-electric UV heating. It is important to note that in our simulations the two-phase ISM is not postulated but a natural result of the physics included in the model.


\subsection{Star formation and feedback}\label{section_stars_feedback}

Star formation and feedback are of crucial importance for the evolution of the galaxy. Our models start out with an initial stellar population described by Eq. (\ref{eq_stellar_density}), but during the simulation  stars keep forming, in the form of stellar particles representing stellar populations. Star formation is based on a Jeans criterion, where a region is considered unstable to star formation when the local Jeans mass is smaller than the mass of a typical molecular cloud, $\mathrm{M_{ref}}$. The rate of star formation is set to scale with the local free fall time. A gas particle that is forming stars converts a fraction $\epsilon_{\mathrm{sf}}$ of its mass into stars, which sets a minimum to the star formation efficiency. However, the actual star formation efficiency is determined by feedback effects from the stars and cooling properties of the gas. This relatively simple recipe for star formation reproduces the Schmidt power law dependence of the SFR on gas density in good agreement with observations, without actually imposing it.

We assume that stars form according to a universal initial mass function (IMF). In this work we take a Salpeter IMF with a lower mass cutoff of 0.1 $\mathrm{M_{\odot}}$. This appears to be a good choice for star formation in the local Universe, however, at high redshift star formation might be described more accurately by a top-heavy IMF. In that case we would be underestimating both the number of ionising photons that are produced by the stars in our simulations and the number of supernova explosions. However, a top-heavy IMF is expected to occur only in very low metallicity gas and in this work we assume that the gas in the galaxies has already been enriched with metals to such an extent that the star formation process is similar to that in the local Universe, leading to an IMF that is comparable to the IMF at low redshift. Recent simulations indicate that this is  indeed the case at redshift 11 \citep{Maio:2010p3144,Maio:2010p3598,Wise:2010p3587}. In addition, previous work on escape fractions by \citet{Wise:2009p1755} showed that the difference in escape fraction between galaxies with a top-heavy and a Salpeter IMF is at most 75\% when the strength of the supernova feedback is kept constant at the level of the top-heavy IMF.  However, self consistent adjustment of the supernova feedback strength with the IMF would result in a larger difference between standard and top-heavy models. We plan to study the effect of top heavy IMFs in future work. 

In our models we incorporate two types of stellar feedback. The far-UV radiation from the stars heats the gas, which is taken into account in the code as part of our model of the ISM. The far-UV luminosity of the stellar particles are calculated using an updated version of the \citet{Bruzual:1993p3007} population synthesis models. Furthermore, stars inject energy into the gas by stellar winds, supernovae and the expansion of \Hii regions. These three effects constitute the mechanical luminosity of the stellar particles. The mechanical energy is dominated by the energy output by supernovae, which shape the ISM surrounding the star formation sites. In our models we assume that stars heavier than 8 $\mathrm{M_{\odot}}$ explode as type {\sc II} supernovae with energies of $10^{51}$ ergs. Combined with the IMF this sets the energy injection by supernovae at $9 \cdot 10^{48}$ ergs per solar mass of total stars formed. Not all of this energy goes into the mechanical feedback, as part of the energy is radiated away in thin, dense shells surrounding the bubbles created by the supernovae, that are unresolved in our simulations. We assume that 10 \% of the total supernova energy goes into mechanical feedback on the gas \citep{Wheeler:1980p3813,Silich:1996p3814}. The feedback is incorporated in the code with the use of pressure particles \citep{Pelupessy:2004p2998,Pelupessy:2005p3001}. Note that our choices for a Salpeter IMF and resulting supernova feedback may have an impact on the resulting escape fractions. We will discuss the possible implications in Sect.~\ref{section_discussion}.


\subsection{Radiative transfer}

We determine the amount of radiation that is able to escape from the galaxies by post-processing the galaxy models with radiative transfer calculations. We thereby neglect the hydrodynamic influence of the radiation on the gas. We do not expect this to have a very large influence on the results, as the energy budget of the expanding \Hii regions is already part of the stellar feedback model (see Sect.~\ref{section_stars_feedback}). Furthermore, \citet{Gnedin:2008p2449} found that the direct coupling of radiation and gas does not change the escape fraction significantly.

In this work we only consider the ionisation of hydrogen atoms, for which photons with energy above 13.6 eV are responsible. The intensity and spectrum of ionising photons produced by the sources in the simulation is calculated using an updated version of the \citet{Bruzual:1993p3007} population synthesis models. The radiative transfer calculations are performed using the {\sc SimpleX2} algorithm \citep{Paardekooper:2010p2772}. {\sc SimpleX} utilises an unstructured grid to transport photons on. The grid is constructed such that regions where the optical depth of the medium is higher are sampled with higher resolution, resulting in an efficient method in which the computational cost is independent of the number of sources in the simulation. The latter property is important given the large number of sources present in the galaxies. 

Each output of the hydrodynamic simulation is post-processed with radiative transfer until the ionisation state and the escape fraction reach an equilibrium value. This typically happens within 2 to 3 Myr and always within 5 Myr. Because the gas is initially almost fully neutral, this provides an upper limit on the time it takes before convergence is reached. As this upper limit is well within the time scale at which source evolution takes place (typically 30 Myr), this is another indication that post-processing the simulations with the radiative transfer calculations does not affect our results significantly. 


\subsection{Dust}

In the local Universe, dust is mainly produced by AGB stars \citep{Gehrz:1989p3077,Dwek:1998p3087,Woitke:2006p3086,Hofner:2007p3075,Mattsson:2008p3076}. At higher redshift, they may not contribute as much as the time it takes for stars to reach the AGB becomes comparable to the lifetime of the Universe. However, observations of high-redshift sub-mm galaxies, high-redshift
quasars and GRB afterglows show that dust is present in the early Universe \citep{Maiolino:2004p3070,Perley:2009p3071,Michaowski:2010p3072}. This dust may have been produced in supernova explosions \citep{Todini:2001p3080,Morgan:2003p3081,Nozawa:2003p3082,Hirashita:2005p3083,Dwek:2007p3078}, but recent work shows that the contribution by AGB stars can't be neglected even at high redshift, as the AGB stars begin to dominate dust production over SNe as early as 150-500 Myr after the onset of star formation \citep{Valiante:2009p3084}.

The presence of dust in the early Universe makes it important to take absorption by dust into account in our calculations of escape fractions. To this end, we use the prescription proposed by
\citet{Gnedin:2008p2449}. This model is based on the observed extinction curves of the Small and Large Magellanic Clouds. Using these measurements, we can express the cross section of dust as an effective cross section per hydrogen atom and thus avoid any assumption about the size distribution and shape of the dust grains. The assumption that goes into this prescription is that the dust at high redshift has similar properties as the dust in the SMC and LMC, which is not obvious since at high redshift the sources of dust might be different. 

\begin{figure} 
  \centering
  \includegraphics[width=9cm]{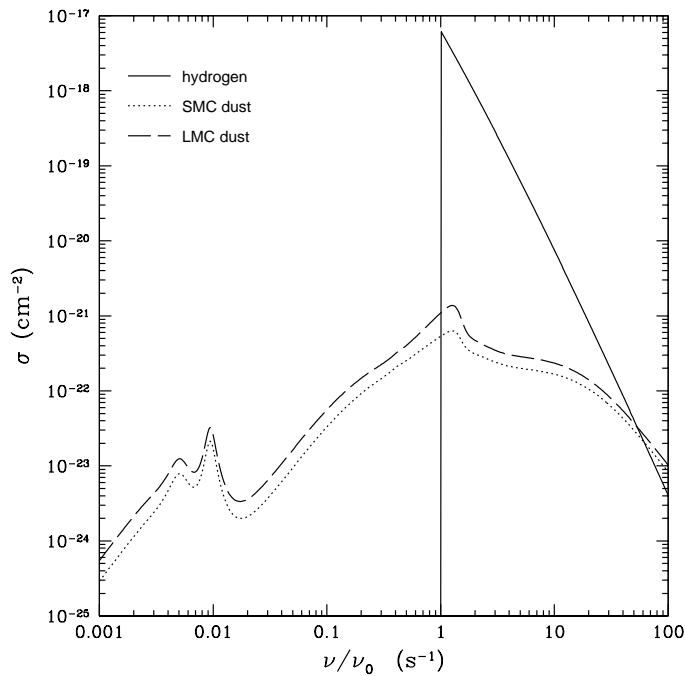}
     \caption{Cross section of hydrogen (solid), SMC dust (dotted) and LMC dust (dashed) as function of frequency normalised to the frequency of the Lyman limit.}
        \label{fig_cross_sections}
\end{figure}

The optical depth due to dust is
\begin{equation} \label{eq_tau_dust} 
  \tau_{\mathrm{d}} = N_{\mathrm{d}} \sigma_{\mathrm{d}}, 
\end{equation}
where $N_{\mathrm{d}}$ is the dust column density and $\sigma_{\mathrm{d}}$ is the dust cross section. For the latter we use the fits from \citet{Gnedin:2008p2449} who updated the analytic fits provided by \citet{Pei:1992p2708}. These cross sections are shown in Fig.~\ref{fig_cross_sections}.
The difference in cross section between SMC and LMC dust is mainly due to the higher metallicity of the latter. Both dust cross sections are well below the photoionisation cross section for hydrogen in almost the entire frequency range above the Lyman limit. 

Since we expressed the dust cross section as an effective cross section per hydrogen atom, we should use the hydrogen number density  instead of the dust number density in Eq.(\ref{eq_tau_dust}). However, in order to account for a metallicity different from that in the SMC and LMC we scale the hydrogen number density with the metallicity. Furthermore, to account for the destruction of dust we consider two cases. The first case is that dust is not destroyed at all, so dust scales with the total hydrogen number density:
\begin{equation} 
  n_{\mathrm{d}} = n_{\mathrm{H}} \frac{Z}{Z_{0}}.
\end{equation}
In the second case the dust is completely sublimated in the ionised regions, so dust scales with the neutral hydrogen number density:
\begin{equation} 
  n_{\mathrm{d}} = (1 - \chi) n_{\mathrm{H}} \frac{Z}{Z_{0}}.
\end{equation}
Here, $\chi$ is the ionised fraction, $Z$ is the metallicity and $Z_{0}$ the reference metallicity of the SMC and LMC relative to solar metallicity. We use $Z_{0,\mathrm{SMC}} = 0.25$ and $Z_{0,\mathrm{LMC}} = 0.5$ \citep{Welty:1997p2784,Welty:1999p2785}. In this work, we will primarily use the SMC dust model as its lower metallicity and relatively young stellar population are most likely a closer match to high-redshift dust extinction than the LMC. However, tests show that for our calculations the difference in escape fractions between the two dust models is less than 1\%.


\subsection{Calculation of the escape fraction}

We calculate the escape fraction by comparing the number of ionising photons that are produced by the stars in a time step to the number of photons that travel beyond $r_{200}$:
\begin{equation} 
f_{\mathrm{esc}} (t) = \frac{N_{\mathrm{phot}}(r > r_{200}, t) }{N_{\mathrm{emitted}}(t)}.
\end{equation}
This is the absolute escape fraction of ionising photons at a certain time $t$. Note that there is a small delay between the moment a photon is emitted and the moment it escapes. This delay is equal to the light travel time from a source to $r_{200}$, which is, depending on the mass of the galaxy, between 11 and 89 kyr in our simulations. This is much shorter than the dynamical time scale of star formation, so this poses no problem.


\section{Results}\label{section_results}

In this section we present the results of post-processing the galaxy simulations with radiative transfer. We will first focus on general trends in the escape of radiation from galaxies, after which we will discuss the influence of various physical parameters on the star formation rate and the escape fraction.


\subsection{Galaxy morphologies}

\begin{figure} 
  \centering
  \includegraphics[width=8cm]{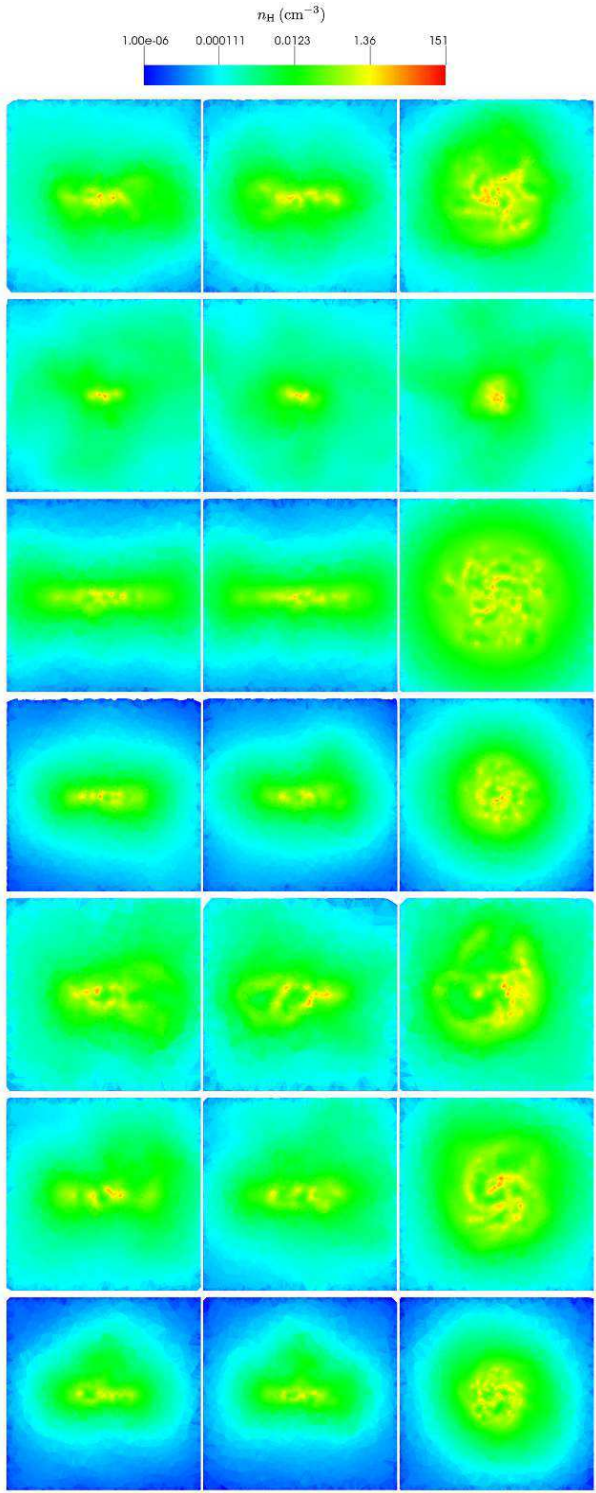}
     \caption{Slices through the x-axis (left), y-axis (centre) and z-axis (right) of the gas density distribution of the $10^{8} M_{\odot}$ galaxies. From top to bottom galaxies 1 to 7 are shown. The physical parameters of these galaxies are listed in Table \ref{table_models}. }
        \label{fig_densities_1e8}
\end{figure}

\begin{figure} 
  \centering
  \includegraphics[width=8cm]{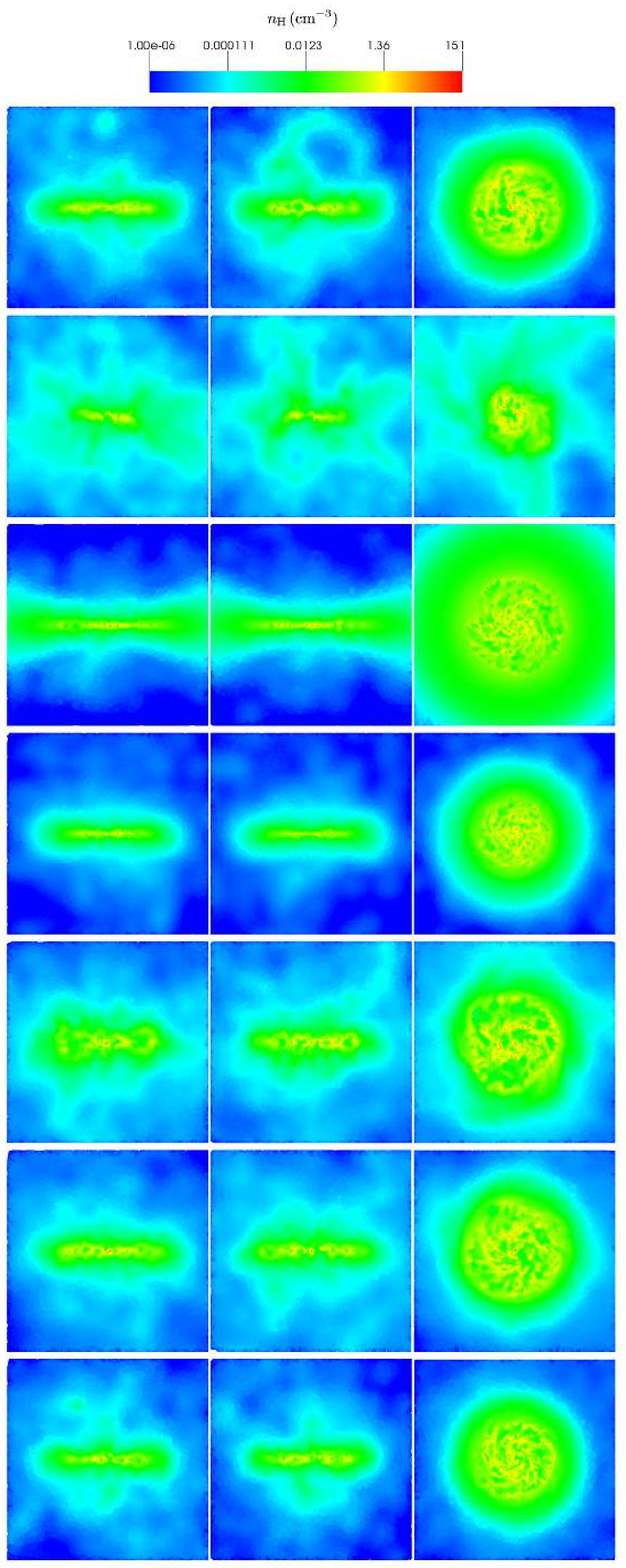}
     \caption{Slices through the x-axis (left), y-axis (centre) and z-axis (right) of the gas density distribution of the $10^{9} M_{\odot}$ galaxies. From top to bottom galaxies 8 to 14 are shown. The physical parameters of these galaxies are listed in Table \ref{table_models}.}
        \label{fig_densities_1e9}
\end{figure}

The density distributions of the $10^{8} M_{\odot}$ and $10^{9} M_{\odot}$ galaxies after 550 Myr are shown in Figs.~\ref{fig_densities_1e8} and \ref{fig_densities_1e9}.
Even though initially all models consisted of stars confined inside a disc of gas, physical processes within the galaxies quickly cause deviations from these initial conditions.  The
morphologies of the $10^{8} M_{\odot}$ and $10^{9} M_{\odot}$ galaxies follow similar trends as the physical parameters change. The more massive galaxies show more substructure and higher peak densities, but the overall morphology is comparable to the lower mass galaxies for the same set of parameters. 

Figs.~\ref{fig_densities_1e8} and \ref{fig_densities_1e9} show that different choices for the parameters of our galaxy models give rise to different morphologies. The spin parameter has the most pronounced effect. A high value for the spin parameter results in a gas distribution that is confined in a disc, where density peaks occur out to large radii. A lower spin parameter results in a highly irregular density distribution where supernova explosions can easily expel gas from the dark matter halo and high densities are found only in the centre of the galaxy. The effect of the initial gas fraction and formation redshift on the morphologies of the galaxies is smaller. A high initial gas fraction leads to dense clumps that are more extended and reach out to larger radii, while a low initial gas fraction has the opposite effect. Apart from the compactness of the halo, there is no significant change in morphology as result of changing the formation redshift of the galaxy. 

The high density knots in the galaxies are the main sites of star formation. Column densities in these regions are so high that ionising photons are trapped inside. There are two ways in which the ionising radiation produced by the stars might escape from these high density regions. Stars can migrate from their formation sites towards lower density environments, so that the ionising photons are no longer trapped and can escape more easily. However, in the time that this migration takes place the massive sources that produce most ionising radiation have ceased to exist. Source migration alone can therefore not provide an efficient mechanism for the bulk of ionising radiation to escape. An alternative way in which the column density in the vicinity of the stars can be lowered is by supernovae. Supernova explosions can blow holes in the gas, thus creating channels through which the radiation can travel. In this case, radiation can escape only in directions where supernovae have created gaps in the gas density distribution, making the escape of ionising radiation highly inhomogeneous. 

These two ways in which ionising radiation can escape from the high density environments in which star formation takes place indicate that the timing of the star formation is very important. As the scales at which individual stars form cannot be resolved in current simulations, stellar particles represent a population of stars. The time at which the bulk of UV radiation is produced and the energy released by supernovae are both governed by subgrid physics. If young stars are still abundant when supernova explosions have created gaps in the high density gas distribution many UV photons will escape, but if at that time the young stars are no longer present there will not be many photons escaping. This means that the subgrid physics of star formation and feedback can play an important role in the determination of the escape fraction of ionising radiation. 

If supernova explosions are indeed the catalyst of radiation escaping from the galaxies, we expect a correlation between the star formation rate in individual galaxies and the fraction of radiation escaping. This correlation is not due to the increasing number of ionising photons that are produced when star formation rates are higher, because without channels to escape from all photons will be trapped. However, a higher star formation rate also implies more supernova explosions, which in turn results in a higher chance that channels through which radiation can escape will be created. We therefore expect high escape fractions in galaxies with high star formation rates. This correlation between star formation and escape fraction was reported earlier by \citet{Wise:2009p1755}, although this effect was particularly strong in their simulations due to their very high supernova feedback strength. In order to study the interplay between star formation rate and escape fractions in more detail, we will first discuss the influence of the physical parameters of the galaxy on the star formation rate before turning our attention to the escape fractions. 


\subsection{Star formation rates}

\begin{figure*} 
  \centering
  \includegraphics[width=\textwidth]{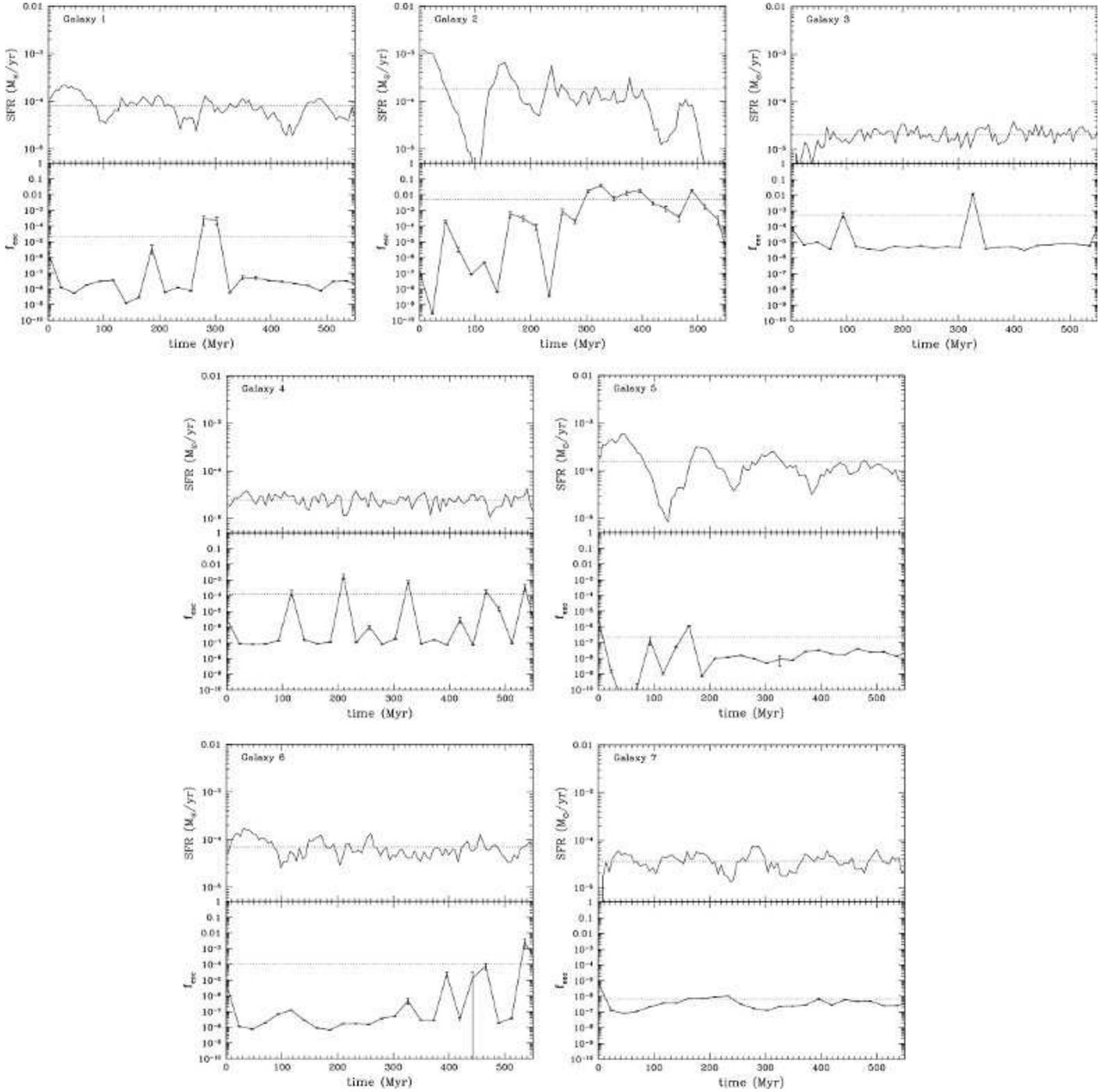}
     \caption{The star formation rate and escape fraction as function of time for the $10^{8} M_{\odot}$ galaxies. The physical parameters of these galaxies are listed in Table \ref{table_models}. The dotted lines represent the time-averaged values.}
        \label{fig_escape_fractions_1e8}
\end{figure*}

\begin{figure*} 
  \centering
  \includegraphics[width=\textwidth]{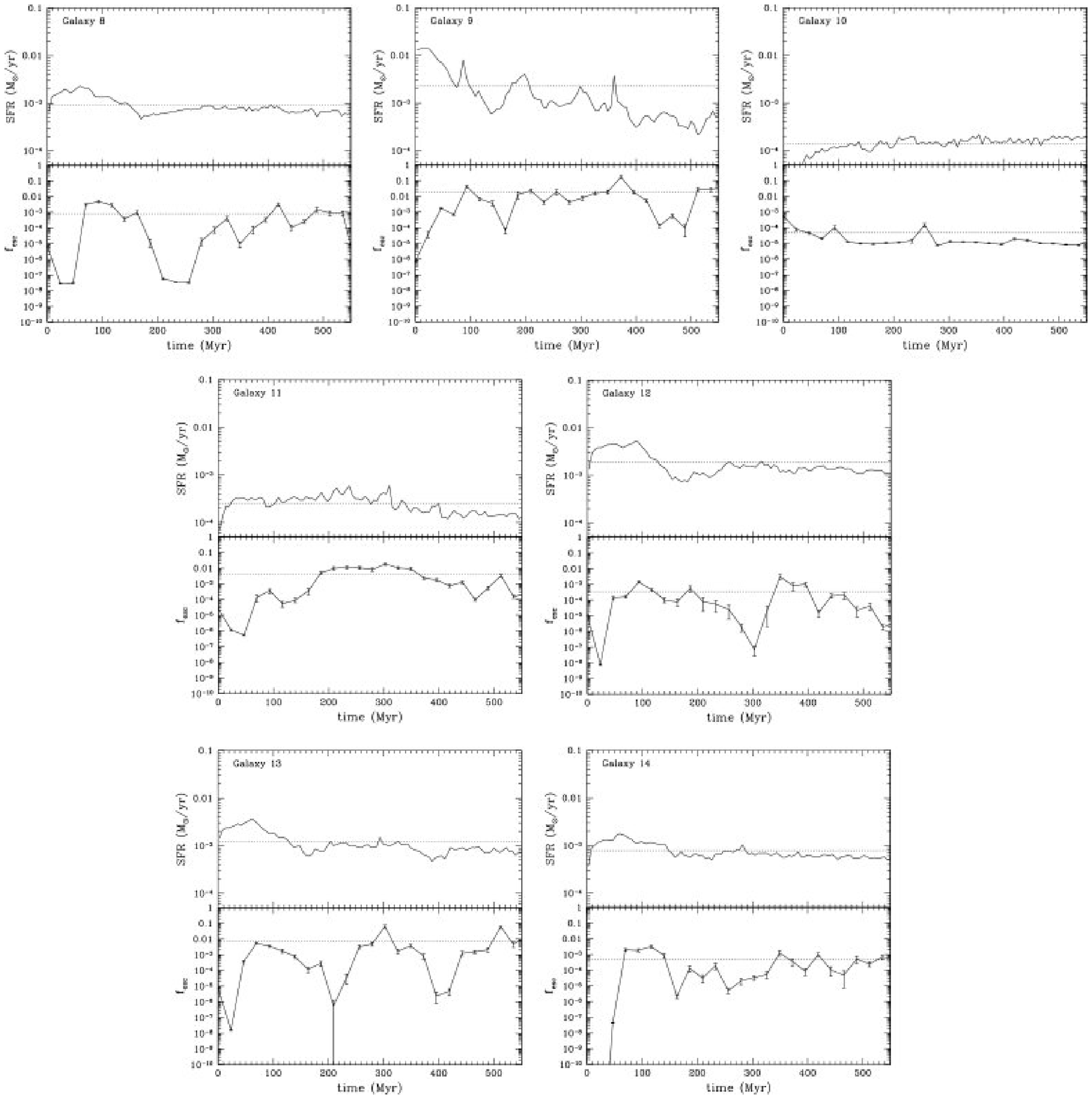}
     \caption{The star formation rate and escape fraction as function of time for the $10^{9} M_{\odot}$ galaxies. The physical parameters of these galaxies are listed in Table \ref{table_models}. The dotted lines represent the time-averaged values.}
        \label{fig_escape_fractions_1e9}
\end{figure*}

The star formation rate as a function of time in our model galaxies is shown in the top panels of Figs.~\ref{fig_escape_fractions_1e8} and \ref{fig_escape_fractions_1e9}. 
The star formation rates averaged over the lifetimes of the galaxies range between $10^{-5}$ and $10^{-3} M_{\odot}\mbox{ yr}^{-1}$ for the $10^{8} M_{\odot}$ galaxies and between $10^{-4}$ and $10^{-2} M_{\odot}\mbox{ yr}^{-1}$ for the $10^{9} M_{\odot}$ galaxies. The higher star formation rates in the higher mass galaxies are due simply to the fact that there is more gas available in these galaxies. Most galaxies show a quiet evolution with a relatively constant star formation rate. Exceptions are the galaxies with low spin parameter or high gas fraction, which show evidence of periodic bursts of star formation. In these cases the star formation rate deviates almost one order of magnitude from the mean. 

The star formation rates we find are consistent with what has been found previously for local dwarf galaxies, both in observations (e.g. \citet{vanZee:2001p3235,DohmPalmer:1998p3282,DohmPalmer:2002p3283}) and simulations (e.g. \citet{Andersen:2000p3284,Mayer:2001p3286,Pasetto:2003p3285}), but are considerably lower than found in previous work by \citet{Wise:2009p1755} and \citet{Razoumov:2010p1840} on high-redshift dwarf galaxies. This is not surprising as those two studies considered dwarf galaxies in which starbursts prevented the gas from settling into a disc. Our models with low spin parameters have irregular morphologies and show the same burst-like behaviour of star formation, with the peak star formation rate comparable to what was found by \citet{Wise:2009p1755} and \citet{Razoumov:2010p1840}.

\begin{figure*} 
  \centering
  \includegraphics[width=\textwidth]{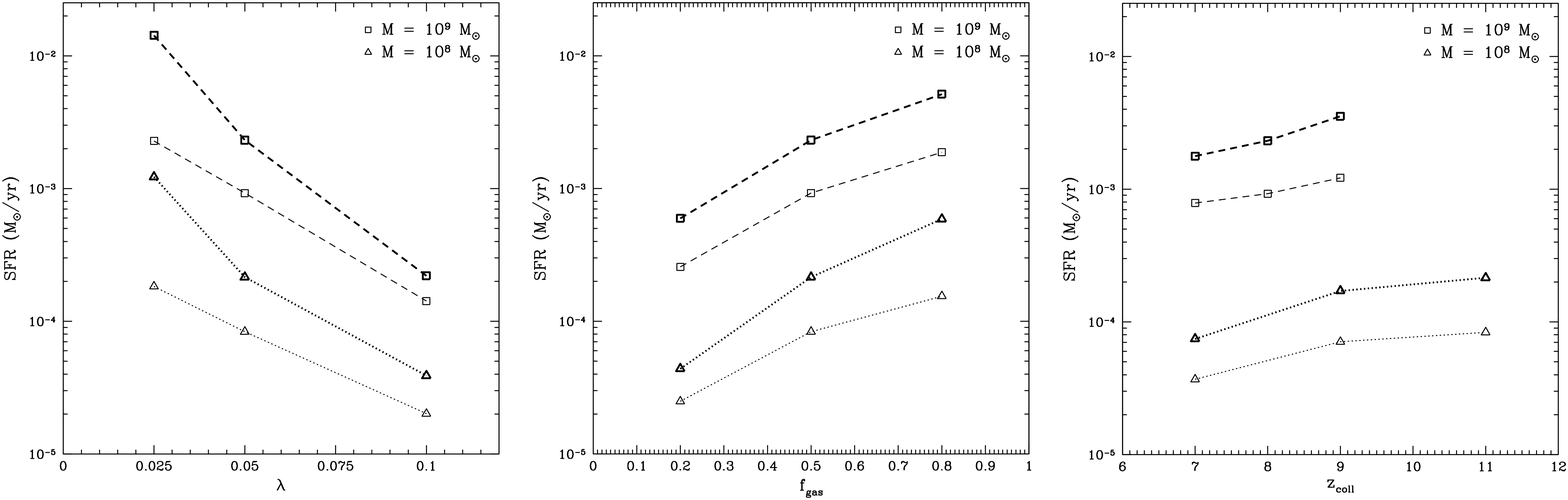}
     \caption{\textit{left:} Mean star formation rate (thin lines) and maximum star formation rate (thick lines) as function of spin parameter $\lambda$. All galaxies have an initial gas fraction $f_{\mathrm{gas}} = 0.5$ and the formation redshift is $z=11$ for the $10^{8} M_{\odot}$ galaxies and $z=8$ for the $10^{9} M_{\odot}$ galaxies. \textit{centre:} Mean star formation rate (thin lines) and maximum star formation rate (thick lines) as function of the initial gas fraction. All galaxies have a spin parameter of 0.05 and the formation redshift is $z=11$ for the $10^{8} M_{\odot}$ galaxies and $z=8$ for the $10^{9} M_{\odot}$ galaxies. \textit{right:} Mean star formation rate (thin lines) and maximum star formation rate (thick lines) as function of formation redshift. All galaxies have a spin parameter of 0.05 and an initial gas fraction of 0.5.}
        \label{fig_SFR_param}
\end{figure*}

Fig.~\ref{fig_SFR_param} shows the dependence of the mean and maximum star formation rate over the lifetime of the galaxy on the initial conditions of the galaxy. 
The left hand side of this figure shows the dependence of the star formation rate on the spin parameter of the galaxy. All galaxies in this plot have an initial gas fraction of 0.5, while the formation redshift is $z=11$ for the $10^{8} M_{\odot}$ galaxies and $z=8$ for the $10^{9} M_{\odot}$ galaxies. The spin parameter ranges from 0.025 to 0.1. The resulting star formation rates vary over an order of magnitude, from $ \sim 2 \cdot 10^{-5}$ to $ \sim 2 \cdot 10^{-4} \, \, M_{\odot} \mbox{ yr}^{-1}$ for the $10^{8} M_{\odot}$ galaxies and from $ \sim 10^{-4}$ to $ \sim 2 \cdot 10^{-3} \, \, M_{\odot} \mbox{ yr}^{-1}$ for the $10^{9} M_{\odot}$ galaxies, with star formation rate decreasing with increasing spin parameter. The trend is present both in the high and the low mass galaxies. A low spin parameter results in gas flowing towards the centre of the halo, where the majority of star formation is taking place, while for higher spin parameters the star formation is spread out in the disk. The models with low spin parameters show periodic bursts of star formation, which results in a maximum star formation rate that deviates almost an order of magnitude from the mean.

The centre plot in Fig.~\ref{fig_SFR_param} shows the dependence of the mean star formation rate on the initial gas fraction in the galaxy. All galaxies in the figure have a spin parameter of $\lambda = 0.05$ and the same formation redshifts as the left hand plot. Again a clear trend is visible, with a higher star formation rate for higher initial gas fractions. As with the spin parameter, the star formation rate varies over an order of magnitude, ranging from $ \sim 2 \cdot 10^{-5}$ to $ \sim 2 \cdot 10^{-4} \, \, M_{\odot} \mbox{ yr}^{-1}$ for the $10^{8} M_{\odot}$ galaxies and from $ \sim 10^{-4}$ to $ \sim 2 \cdot 10^{-3} \, \, M_{\odot} \mbox{ yr}^{-1}$ for the $10^{9} M_{\odot}$ galaxies. The star formation rate increases with increasing gas fraction. This can be explained by the fact that a higher initial gas fraction means that there is more gas available for star formation at later stages in the lifetime of the galaxy.  

Finally, the right hand side of Fig.~\ref{fig_SFR_param} shows the dependence of the mean star formation rate on the formation redshift of the galaxy. All galaxies in this figure have a spin parameter of $\lambda = 0.05$ and an initial gas fraction of 0.5. In our models the formation redshift determines the compactness of the dark matter halo. The higher central densities of the dark matter halo at higher redshift makes it easier to form stars. However, the influence of the formation redshift is small compared to the spin parameter and initial gas fraction. In contrast to these two parameters, the mean star formation rate varies no more than 40\% over the entire redshift range, from $ \sim 3 \cdot 10^{-5}$ to $ \sim 9 \cdot 10^{-5} \, \, M_{\odot} \mbox{ yr}^{-1}$ for the $10^{8} M_{\odot}$ galaxies and from $ \sim 8 \cdot 10^{-4}$ to $ \sim 10^{-3} \, \, M_{\odot} \mbox{ yr}^{-1}$ for the $10^{9} M_{\odot}$ galaxies.


\subsection{Escape fractions} \label{section_results_escape}

Given the strong correlation between star formation rate and the escape fraction of ionising radiation found in previous studies, we would expect that the escape fraction follows the same trend as the star formation rate. In the individual galaxies in Figs.~\ref{fig_escape_fractions_1e8} and \ref{fig_escape_fractions_1e9} the escape fraction indeed follows the trends in star formation rate as a function of time. The effect is most pronounced in the galaxies with low spin parameter, in which the star formation happens in bursts. Because the gas in these galaxies is only weakly bound to the disc as a result of the low spin parameter, the effect of supernova explosions is more severe than in the other galaxies. The figures show that if the spin parameter is low, a peak in the star formation rate results in a peak in the escape fraction. This is not due to the increase in number of ionising photons, but to the increase in the number of supernova explosions that create channels through which the radiation can escape. The other galaxies also show some evidence of this effect, albeit less pronounced.

The escape fraction shows a highly irregular behaviour in all galaxies, with escape fractions changing over 3 and sometimes even 10 orders of magnitude during the simulation time. In galaxies with a highly variable star formation rate this is caused by the correlation between star formation rate and escape fraction. However, galaxies in which the star formation rate does not change significantly over time show highly variable escape fractions as well. Even though the number of supernova explosions in these galaxies is relatively constant over time, not all these explosions result in escaping radiation. A peak in the escape fraction only occurs when a supernova creates a bubble large enough to provide a channel for the ionising radiation. This results in variations in the escape fraction of 3 orders of magnitude even when the star formation rate is relatively constant over the simulation time. However, these variations are smaller than the variations in escape fraction that occur when the star formation happens in bursts, in which case the escape fraction can change up to 6 orders of magnitude following the variation of the star formation rate with time. 

\begin{figure*} 
  \centering
  \includegraphics[width=\textwidth]{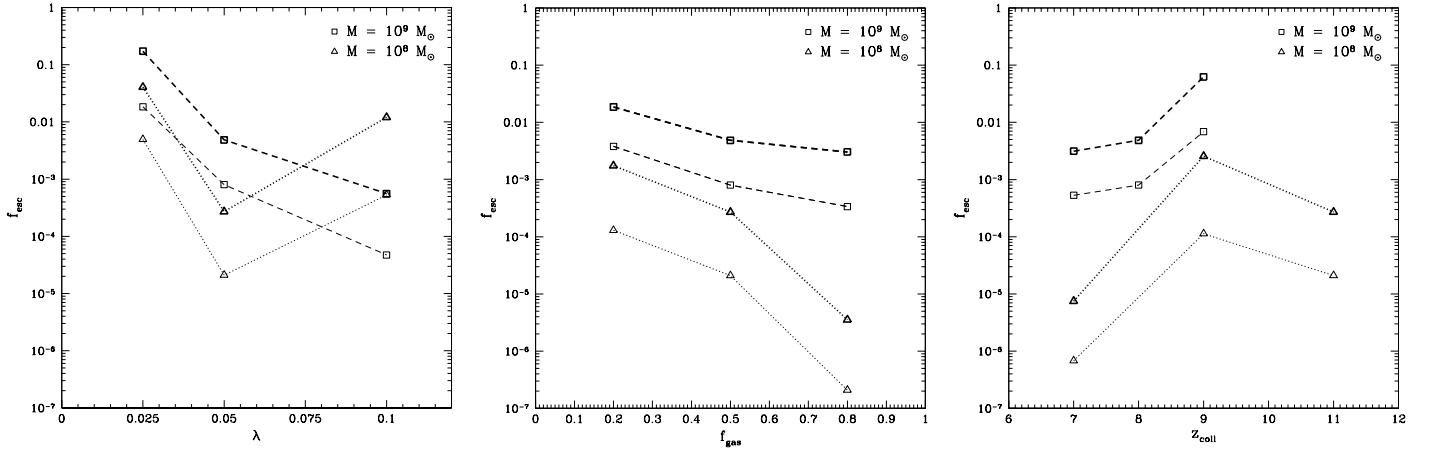}
     \caption{\textit{left:} Mean escape fraction (thin lines) and maximum escape fraction (thick lines) as function of spin parameter $\lambda$. All galaxies have an initial gas fraction $f_{\mathrm{gas}} = 0.5$ and the formation redshift is $z=11$ for the $10^{8} M_{\odot}$ galaxies and $z=8$ for the $10^{9} M_{\odot}$ galaxies. \textit{centre:} Mean escape fraction (thin lines) and maximum escape fraction (thick lines) as function of the initial gas fraction. All galaxies have a spin parameter of 0.05 and the formation redshift is $z=11$ for the $10^{8} M_{\odot}$ galaxies and $z=8$ for the $10^{9} M_{\odot}$ galaxies. \textit{right:} Mean escape fraction (thin lines) and maximum escape fraction (thick lines) as function of formation redshift. All galaxies have a spin parameter of 0.05 and an initial gas fraction of 0.5.}
        \label{fig_escape_param}
\end{figure*}

The large variability of the escape fraction over time in individual galaxies makes it hard to find global trends with changing physical parameters of the galaxies, as the deviation from the time averaged escape fraction is quite large. This is due to the fact that the main feature determining the escape fraction, the occurrence of low density bubbles caused by supernova explosions, is only indirectly related to the galaxy parameters. Despite the variability in the mean escape fraction of individual galaxies it is possible to observe some general trends as the physical parameters change. In the individual galaxies, the escape fraction generally follows the star formation rate. However, the mean escape fractions do not always scale the same way as the mean star formation rates with respect to the galaxy model parameters.  Fig.~\ref{fig_escape_param} shows the mean and maximum escape fraction as function of spin parameter, initial gas fraction and formation redshift for the same galaxies as plotted in Fig.~\ref{fig_SFR_param}. 
The left hand side of this figure shows that the escape fraction does generally follow the same trend as the star formation rate, supporting the notion that escape fraction and star formation rate are coupled. The general trend is a declining escape fraction with increasing spin parameter, with the escape fraction varying over almost 3 orders of magnitude, ranging from $ \sim 10^{-5}$ to $ \sim 10^{-2}$ for the $10^{8} M_{\odot}$ galaxies and from $ \sim 5 \cdot 10^{-5}$ to $ \sim 5 \cdot 10^{-2}$ for the $10^{9} M_{\odot}$ galaxies. The exception is the $10^{8} M_{\odot}$ galaxy with $\lambda = 0.1$ that shows an increase in escape fraction. However, the mean value of the escape fraction of this particular galaxy is dominated by only a few peaks in the escape fraction, so in this case the mean is not representative of the majority of the time the galaxy is emitting photons. Overall, the escape fractions show a similar trend with $\lambda$ as the mean star formation rate. A higher spin parameter keeps the galaxy disc-like with the stars trapped in the middle of the disc. In order to escape the radiation has to travel through the disc where the column density is highest, resulting in a low value for the escape fraction. 

In contrast to the spin parameter, the escape fraction shows a different dependence on the initial gas fraction compared to the star formation rate. Where the star formation rate increases with gas fraction, the escape fraction declines with higher gas fraction, shown in the centre plot of Fig.~\ref{fig_escape_param}. The mean escape fraction ranges from $\sim 10^{-7}$ to $\sim 10^{-4}$ for the $10^{8} M_{\odot}$ galaxies and from $\sim 3 \cdot 10^{-4}$ to $\sim 5 \cdot 10^{-3}$ for the $10^{9} M_{\odot}$ galaxies, with lower escape fraction for galaxies with a high initial gas fraction. As the gas fraction of the galaxy increases, the radiation has to travel through higher neutral gas column densities to escape. This effect is stronger than the higher star formation rate that in principle could lead to a higher escape fraction.

The escape fraction as function of the formation redshift of the galaxy is shown on the right hand side of Fig.~\ref{fig_escape_param}. A higher formation redshift results in a dark matter halo that is more compact, which makes it easier to form stars. This higher star formation rate allows for more escape of radiation, resulting in a higher escape fraction at higher redshift. Escape fractions vary between $\sim 5 \cdot 10^{-7}$ to $\sim 10^{-4}$ for the $10^{8} M_{\odot}$ galaxies and from $\sim 5 \cdot 10^{-4}$ to $\sim 10^{-2}$ for the $10^{9} M_{\odot}$ galaxies. The general trend is an increasing escape fraction at higher redshift. The $10^{8} M_{\odot}$ galaxy with formation redshift 9 deviates from this trend. As we can see in Fig.~\ref{fig_escape_fractions_1e8} the relatively high escape fraction in this galaxy is caused by a few peaks in the escape fraction near the end of the simulation. In the majority of the time that the galaxy is emitting photons the escape fraction is much lower.

\begin{figure} 
  \centering
  \includegraphics[width=9cm]{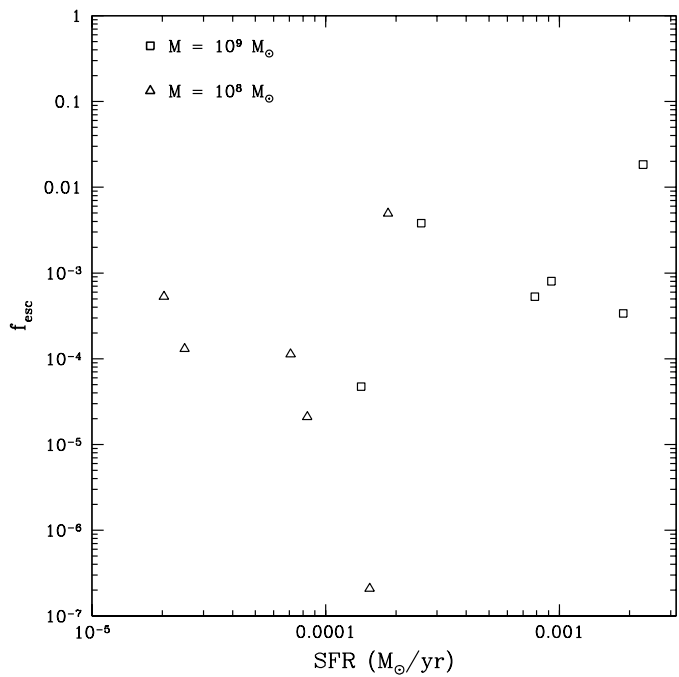}
     \caption{The escape fraction as function of mean star formation rate in the galaxy. }
        \label{fig_escape_SFR}
\end{figure}

Fig.~\ref{fig_escape_SFR} shows the mean escape fractions of all simulated galaxies as function of mean star formation rate. 
This figure shows no evidence of a relation between the mean star formation rate and the mean escape fraction. A high star formation rate results in more feedback on the gas and thus a higher chance of channels being created for radiation to escape, resulting in a high escape fraction. On the other hand, a high star formation rate can also be the result of high gas content in the galaxy, resulting in a higher gas column density that makes it harder for radiation to escape. Therefore, no one-to-one relationship between the mean star formation rate and the mean escape fraction exists in our models.

\begin{figure} 
  \centering
  \includegraphics[width=9cm]{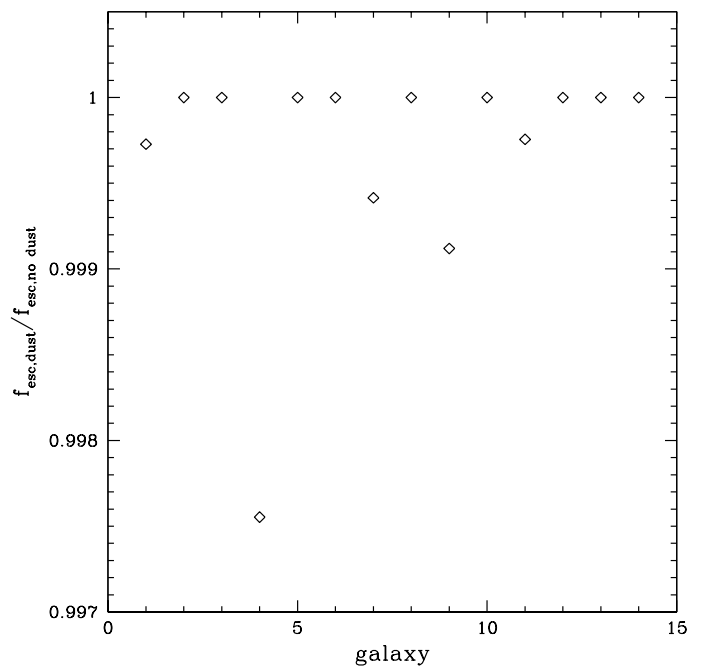}
     \caption{The ratio of escape fraction with and without dust included in the radiative transfer calculation. The dust model is that of the Small Magellanic Cloud, which has a metallicity comparable to that of the model galaxies. To place an upper limit on the effect of dust we assumed no sublimation at all.}
        \label{fig_esc_dust}
\end{figure}

In order to measure the effect of dust on the escape fraction in high-redshift galaxies, we simulated the same set of galaxy models adding dust to the radiative transfer calculation. We used the Small Magellanic Cloud dust model, as the metallicity of the SMC is comparable to the metallicity of our models. To maximise the effect of dust absorption we assumed no dust sublimation at all. We can therefore place an upper limit on the effect of dust on the escape fraction from these galaxies. Fig.~\ref{fig_esc_dust} shows the ratio of escape fraction with and without dust included. 
In all galaxy models the change in escape fraction is below 1\% when dust is included. 

There are several reasons why the influence of dust is so small. First of all, the cross section of the dust is orders of magnitude smaller than the hydrogen cross section at frequencies above the Lyman limit. Therefore only a very small fraction of the radiation is absorbed by dust. Second, we assume the dust number density to follow the gas density. Since radiation preferentially escapes through channels with low gas density there is also not much dust present, which further reduces the effect of dust. Finally, all the model galaxies have a low metallicity of 0.2 solar, which further reduces dust abundance. These results are similar to the results of \citet{Gnedin:2008p2449} and \citet{Razoumov:2010p1840}, who also find a negligible influence of dust on the escape fraction. \citet{Yajima:2009p2719} find a much larger effect of dust, which is due to the very high metallicity in the galaxy they simulate.


\subsection{Observational consequences}

The escape fractions we find from all our model galaxies vary with at least 3 orders of magnitude over the lifetime of the galaxies. The escape of radiation is highly inhomogeneous, with radiation escaping only from low density channels created by supernova explosions. Observational studies therefore need very large samples to get a representative picture of the escape of ionising radiation. If escaping UV radiation from a galaxy is detected this means that a channel through which radiation can escape is present in our line-of-sight. Almost all radiation that is emitted by the visible stars is escaping through this channel. The escape fraction that is inferred from such observations will therefore naturally be very high. However, this is merely an orientation effect, it does not mean that the escape fraction of all stars in the galaxy in all directions is high. Our models show that in directions in which no low density channels have been formed it is impossible for radiation to escape. Furthermore, the radiation from stars that are obscured by high density gas can not be observed and is therefore not taken into account in the determination of the escape fraction. For this reason a high observed escape fraction does not mean that most UV radiation produced in the galaxy is escaping. We expect that high resolution observations of galaxies with high UV escape fractions will show escape from regions in the galaxy with high star formation surface density.


\section{Numerical constraints}\label{section_numerics}

Simulating galaxies and measuring escape fractions is a computationally complicated task. Even when realistic sub-grid models are used for the relevant physical processes in the galaxy itself, to compute the escape fractions a large number of sources need to be included in the radiative transfer calculation. Furthermore, the porosity of the ISM in the galaxy needs to be resolved on such small scales that all channels through which radiation can escape are accounted for. In this section we investigate whether or not it is sufficient to include only massive sources which produce most of the ionising photons in the radiative transfer calculation. We also check whether our models have sufficient resolution to resolve the small scales that are necessary for a realistic estimate of the escape fraction and we study the effect that the timing of the star formation has on the escape fraction.


\subsection{Escape of radiation from luminous sources only}

In simulations of escape fractions, the computationally most demanding task is often the radiative transfer calculation. Galaxies consist of millions of stars whose radiation contributes to the total ionising photon budget of the galaxy. Including all these sources is all but impossible if the computation time of the radiative transfer method scales linearly with the number of sources. The {\sc SimpleX} algorithm that was used in this work does not suffer from this drawback, so all stars in the galaxy are included in the radiative transfer calculation. However, it may not be necessary to include all sources, as the bulk of ionising radiation in the galaxies is produced by the most massive stars that are relatively rare.

We have checked whether this assumption is correct by calculating the escape fractions of our galaxy models when only 2-3\% of the total number of sources is included. These sources contribute more than 99\% to the total ionising radiation. This procedure reduces the total number of sources from on average $ \sim 25000$ to $ \sim 600$ in the $10^8 M_{\odot}$ galaxies and from on average $ \sim 240000$ to $ \sim 6000$ in the $10^9 M_{\odot}$ galaxies, making the problem better suited for most radiative transfer methods. 

\begin{figure} 
  \centering
  \includegraphics[width=9cm]{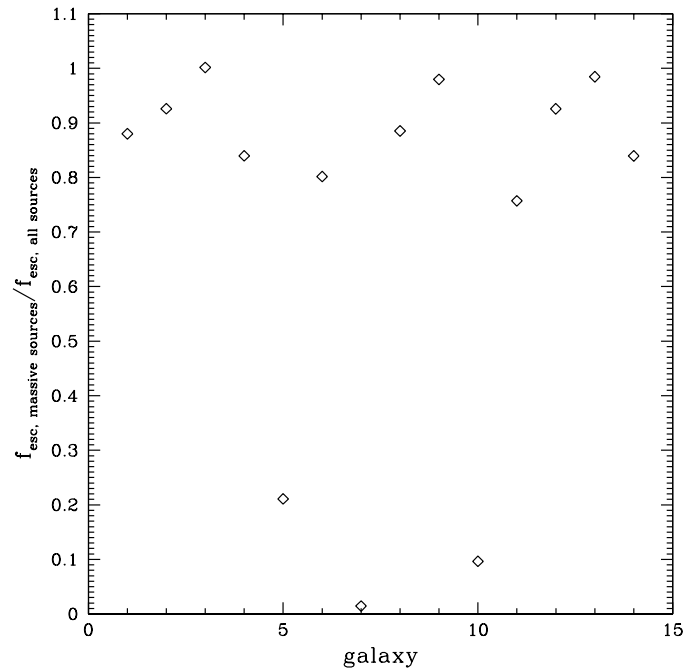}
     \caption{The ratio of escape fraction with only massive sources included and escape fraction with all sources included.}
        \label{fig_escape_sources}
\end{figure}

Fig.~\ref{fig_escape_sources} shows the ratio of escape fraction with only massive sources included and escape fraction with all sources included.
In galaxy 3 the escape fraction is lower when all sources are included. Due to the high angular momentum of this galaxy almost all sources are confined in the high density disc, so the additional ionising flux that the low mass sources contribute cannot escape the galaxy at all, resulting in a lower escape fraction. This galaxy is the only one in which this happens.

The general trend is a higher escape fraction when the low mass sources are included in the simulation. Despite the fact that these less massive sources contribute less than 1\% of the ionising photons, for most galaxies escape fractions with all sources included are 10-20\% higher and in some cases an order of magnitude. This could be an indication that radiation from the low mass sources escapes more easily from the galaxy. Low mass sources live longer than high mass sources and therefore have more time to migrate from their formation site to lower density regions from which the radiation can more easily escape. \citet{Gnedin:2008p2449} found that ionising radiation primarily escapes from sources located on the outside of the disc, which supports this scenario. However, we find that even though this migration does take place it, in general, does not lead to more escaping ionising radiation from the low mass sources themselves. As the sources grow older they produce less ionising photons and in most cases the ionising flux is not high enough for the \Hii region to break out even when the gas column density that surrounds the sources is lower than in the central parts of the galaxy. 

The contribution of the low mass sources to the escape fraction lies in their spatial distribution. Consider the case where only one source resides in the centre of a dense star forming region. The source will have to ionise most of the surrounding gas before radiation can break out. Now suppose that the same luminosity is divided over three sources that are located at the vertices of a triangle in the same star forming region. Radiation will escape more easily in this case because the surface area of the three individual \Hii regions added together is larger than the total area of the \Hii region around the single source. Therefore, the chance that radiation finds a channel through which it can escape is higher.

\begin{figure*} 
  \centering
  \includegraphics[width=\textwidth]{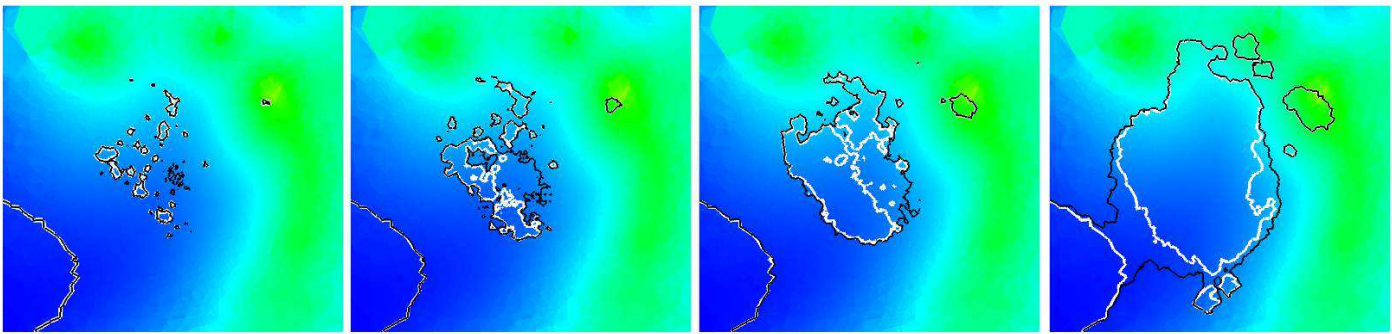}
     \caption{Cuts through the $z = 0.5 \, \, z_{\mathrm{box}}$ coordinate of a snapshot of galaxy 2 with $\lambda = 0.025$, $f_{\mathrm{gas}} = 0.5$ and formation redshift 11 on different times during the radiative transfer calculation. Shown is the density distribution in the central 1.4 kpc of the galaxy. Contours show the regions where the ionised fraction is 0.5, where black contours depict the simulation with all sources included and the white contours the simulation with only massive sources included.}
        \label{fig_as_vs_ms}
\end{figure*}

In our models, the small ionised regions around the low-mass sources provide channels through which the radiation of the massive sources can travel. The larger surface area of all \Hii regions combined makes it easier for the ionising photons to escape. An example of this effect is shown in Fig.~\ref{fig_as_vs_ms}, which shows the central 1.4 kpc of a snapshot of a $10^8 M_{\odot}$ galaxy with $\lambda = 0.025$, $f_{\mathrm{gas}} = 0.5$ and $z = 11$ sliced through the z-axis.
At the beginning of the radiative transfer simulation the low mass sources create very small \Hii regions throughout the disc. The ionising luminosity is not high enough for the \Hii region to break out of the galaxy, but the sources are located outside the star forming region in which the massive sources reside. These tiny ionised bubbles therefore provide channels for the radiation from the more luminous sources to travel, making it easier for all radiation to escape. Fig.~\ref{fig_as_vs_ms} shows that at every time step the surface area of the \Hii regions is significantly larger when low luminosity sources are included. In the final time step of this simulation a bridge between the two separate \Hii regions has been formed, allowing radiation from the inner region to escape. This does not occur in the simulation that excludes the low luminosity sources, resulting in a lower escape fraction. 

These simulations suggests that not only the luminosity of the sources is important for the escape fraction, but also their spatial distribution. Therefore, procedures as source merging and source exclusion may have a large impact on the escape fraction even though the total ionising luminosity of the sources does not change significantly. 


\subsection{Resolution study}

\begin{figure} 
  \centering
  \includegraphics[width=9cm]{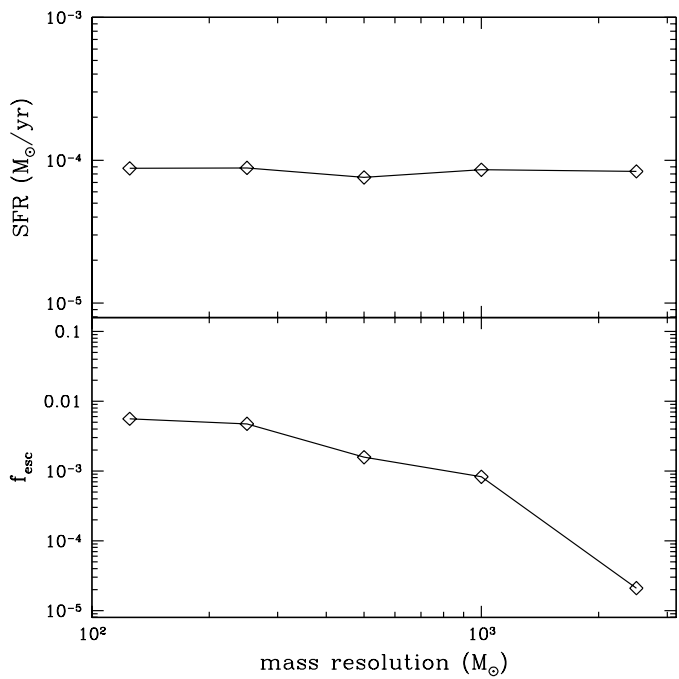}
     \caption{The mean escape fraction and star formation rate over the lifetime of the galaxy as function of mass resolution for galaxy 1. }
        \label{fig_escape_resolution}
\end{figure}

The idea that radiation primarily escapes through low density channels in the gas implies that it's important to resolve the gas up to a scale where the porosity of the ISM is correctly accounted for. We have performed a resolution study to assess whether our simulations have enough resolution to do this by resimulating galaxy number 1 at higher mass resolution. The result of this study is shown in Fig.~\ref{fig_escape_resolution}.
Originally the galaxy was simulated with a mass resolution of 2500 $M_{\odot}$. The top figure shows that the star formation rate in the galaxy does not change when the mass resolution is increased. However, the escape fraction is more than two orders of magnitude higher when the mass resolution is twenty times higher. This is due to the fact that in the high resolution simulations the gas density in the neighbourhood of the sources is better represented. In the highest resolution runs the escape fraction converges to a value of $f_{\mathrm{esc}} \approx 0.005$. In these two runs the average density of the gas surrounding the sources converges as well. We can therefore conclude that the gas density distribution on small scales plays a crucial role in determining the escape fraction. This shows that for accurate estimates, the resolution of both the hydrodynamics and the radiative transfer calculation needs to be high enough to resolve the small scale clumping of the neutral gas. 

Most of the models we ran were chosen to have uniform resolution across galaxy masses and gas fractions, such that the mass resolution was about $1000-2500 \, M_{\odot}$. The above shows that our escape fractions have not converged fully. Also, additional physics may play a role such as feedback by proto-AGN seed blackholes \citep{Pelupessy:2007p3511} or cosmic ray ionizations. The trends with physical parameters that we find for our models should however remain valid, but the escape fractions should be taken only as a lower limit. Note that the value of the escape fraction to which the simulations converge is still below 1\%, which is much lower than found in previous work on similar galaxies. 


\subsection{Local gas clearing}\label{section_timing}

The resolution study in the previous section indicates that the escape fraction depends on the small scale gas density distribution in the vicinity of the sources. This means that the timing of the star formation and the feedback is essential. In our simulations the details of these processes are dealt with in subgrid models, as the resolution is insufficient to resolve the formation of individual stars. Instead the star particles represent a population of stars with masses following the prescribed IMF. We assume that all stars in the population form at the same time and that the massive stars explode as supernovae at the end of their lifetime. This means that the bulk of ionising radiation is emitted at the time that supernova feedback has not been able to create channels in the high density surroundings yet. 

In reality the stars are not likely to form all at the same time, so massive stars will still be forming when stellar feedback has already cleared a fraction of the local gas. Radiation from these sources has therefore a higher chance of escaping, an effect that we have ignored in our simulations. In order to test whether this would affect our results we have performed simulations in which the supernova feedback stays the same but the luminosity of the stars is calculated with a delay after the formation of the star particle. This mimics the behaviour that all stars form with a delay after the onset of stellar feedback, which is of course not realistic. However, it does provide us with an upper limit of the effect of timing in the subgrid physics on the escape fraction and shows whether the escape fraction is constrained primarily by the local gas.

\begin{figure*} 
  \centering
  \includegraphics[width=\textwidth]{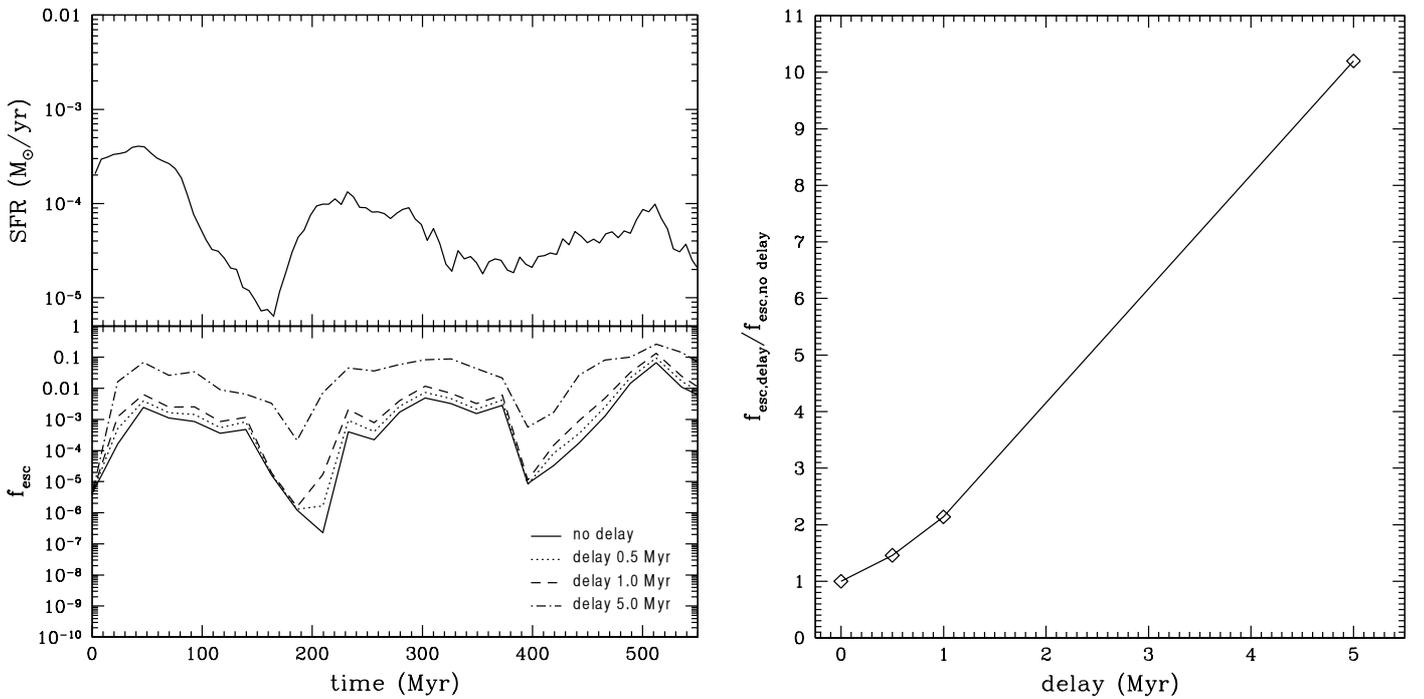}
     \caption{\textit{left:} The star formation rate and escape fraction as function of time for a reference galaxy of $10^8 M_{\odot}$ with $\lambda = 0.05$, $f_{\mathrm{gas}} = 0.5$ and $z=11$ resimulated at mass resolution of 250 $M_{\odot}$. Shown are the escape fraction with no delay in ionising luminosity (solid line), with a delay of 0.5 Myr (dotted line), with a delay of 1 Myr (dashed line) and with a delay of 5 Myr (dashed-dotted line). \textit{right:} Ratio of the escape fraction without delay in ionising luminosity and with delay as function of the delay time.}
        \label{fig_timing}
\end{figure*}

Fig.~\ref{fig_timing} shows the effect that a delay in ionising luminosity has on the escape fraction from our reference galaxy, the $10^8 M_{\odot}$ galaxy with $\lambda = 0.05$, $f_{\mathrm{gas}} = 0.5$ and $z=11$.
The galaxy was resimulated at mass resolution of 250 $M_{\odot}$ to account for the small scale structure of the gas surrounding the sources. Our resolution study shows that the escape fraction has converged at this resolution. We have performed 4 simulations, in which we varied the ionising luminosity of the star particles. In one simulation the UV luminosity was calculated from the formation time of the particles, similar to the simulations presented earlier. In the other simulations the UV luminosity was calculated with a delay of 0.5, 1 and 5 Myr. Stellar particles younger than this delay were excluded from the simulation. Note that these delays are significantly smaller than the typical time period over which the feedback from the stellar particles operates, $\Delta t_{\mathrm{feedback}} = 30 \,\, \mathrm{Myr}$, approximately the lifetime of an $8 \, M_{\odot}$ star. When a delay is applied the peak in ionising luminosity lies at a point where stellar feedback has already had some influence on the gas surrounding the star formation site. The total luminosity of all sources combined is not affected by this procedure if the delay is 1 Myr or less, with total luminosity varying less than 1\%. A delay of 5 Myr results in a $\sim10\%$ increase in total luminosity, due to a higher star formation rate at earlier times.

The right-hand side of Fig.~\ref{fig_timing} shows that the interplay between the stellar feedback and the peak in UV luminosity has a large influence on the escape fraction. If the peak in ionising luminosity is delayed with 0.5 Myr, the escape fraction is 50\% higher, while a delay of 1 Myr results in an escape fraction that is twice as big as in case no delay was imposed. To maximise the effect we have also performed a simulation in which the peak in ionising luminosity lies 5 Myr after the formation of the star particle. This is highly unrealistic as this is much longer than the formation time of the massive stars that are responsible for most of the UV luminosity. It does however give insight into how important the timing is. After 5 Myr the stellar feedback has removed so much gas from the star formation sites that the mean escape fraction is more than 10 times higher. This indicates that the main constraint for the escape of radiation are the local gas complexes that give rise to star formation.

The left-hand side of Fig.~\ref{fig_timing} shows the escape fraction as a function of time for the different simulations. In case of a 5 Myr delay, the supernova feedback has decreased the density around the star formation site so much that the escape fraction is 10 times higher than in the simulation with no delay at all times. The simulations with a 0.5 and 1 Myr delay give better insight into the physical processes at work. Over the lifetime of the galaxy the escape fraction follows the trend set by the star formation, with the escape fraction varying over 5 orders of magnitude. At the times that the escape fraction is low, around 200 and 400 Myr in this simulation, the delay has little effect. These dips coincide with low star formation rates and therefore less stellar feedback. In this case all the sources are deeply embedded in high density regions and it takes time to expel the gas from these regions due to the low star formation rate. Therefore, the 0.5 and 1 Myr delays do not change the escape fraction significantly, although the escape rises earlier from the dip around 200 Myr in case the delays are applied. A stronger influence can be observed when the star formation rate is high and the influence of stellar feedback is more pronounced. In this case a 0.5 (1) Myr delay results in an escape fraction that is 60\% (150\%) higher.

This study shows that the escape of ionising radiation depends crucially on the intricate interplay between star formation and stellar feedback, which clears the gas from the sites of star formation at roughly 50 km/s. Given the strong dependence of the escape fraction on the delay in peak ionising luminosity, this means that it is gas on local ($\ll$ kpc) scale that blocks the majority of radiation. For numerical studies of escape fractions it is therefore essential to model this accurately. 


\section{Implications for reionisation models}\label{section_implications}

Extrapolation of the luminosity function inferred from observations of galaxies during the epoch of reionisation suggest that galaxies must have high escape fractions ($ \sim 20 - 60\%$) to account for the photons needed for reionising the Universe \citep{Labbe:2010p2967,Bouwens:2010p3250}, although these results are highly dependent on the abundance of low mass galaxies at high redshift. It may seem at first that our results indicate that dwarf galaxies do not contribute significantly to reionisation in their quiescent phase, as the escape fractions we find in this work are considerably lower. However, the uncertainties in the measurements of the faint-end-slope during the epoch of reionisation are very large, making conclusive statements in this direction premature.

\begin{figure} 
  \centering
  \includegraphics[width=9cm]{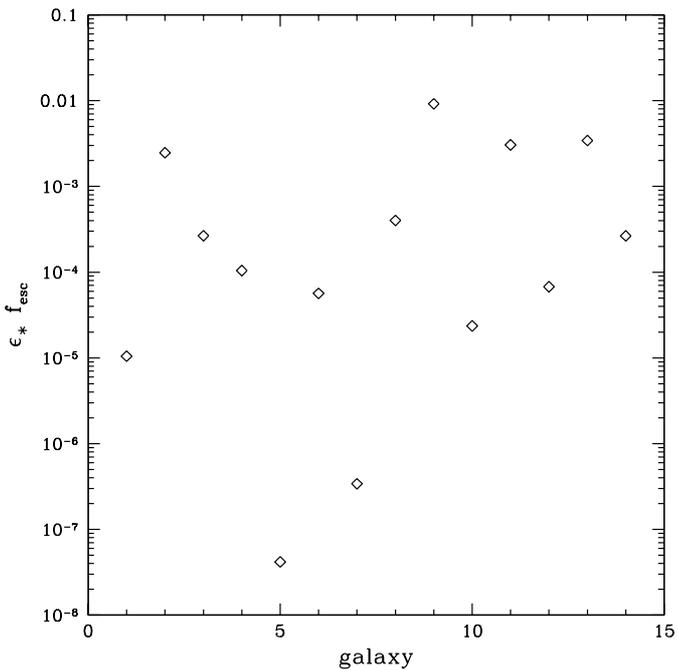}
     \caption{The efficiency of stars in the galaxy to contribute to cosmic reionisation, the parameter used in most semi-analytic models, for our model galaxies. The physical parameters of these galaxies are listed in Table \ref{table_models}.}
        \label{fig_reion}
\end{figure}

We can obtain a more reliable estimate of the contribution of high-redshift dwarf galaxies to cosmic reionisation by comparing our models to the outcome of semi-analytic models of reionisation. In semi-analytic models the relevant parameter for the contribution of haloes to cosmic reionisation is the efficiency of stars $\epsilon = \epsilon_{\ast} f_{\mathrm{esc}}$, where $\epsilon_{\ast}$ is the fraction of baryonic mass in the galaxy converted into stars \citep{Choudhury:2008p2944,Srbinovsky:2010p2800}. This parameter determines how efficient the stars in a galaxy are in reionising the Universe. In Fig.~\ref{fig_reion} we show the mean $\epsilon$ over the computation time for all our model galaxies.
We have used the initial gas fraction $f_{\mathrm{gas}}$ in the galaxy to compute $\epsilon_{\ast}$, thus we are slightly underestimating $\epsilon$ at later times, because stars continue forming. We find that $\epsilon$ ranges from $ \sim 5 \cdot 10^{-8}$ to $ \sim 0.003$ for the $10^{8} M_{\odot}$ galaxies and from $ \sim 3 \cdot 10^{-5}$ to $ \sim 0.01$ for the $10^{9} M_{\odot}$ galaxies. The galaxies that are most efficient contributors to reionisation are those with low spin parameter, with $\epsilon = 0.0025$ and $\epsilon = 0.0092$ for the $10^{8} M_{\odot}$ and $10^{9} M_{\odot}$ galaxy respectively. The galaxies with low gas fraction and high formation redshift show high efficiencies as well, of order $\sim 0.003$ for the $10^{9} M_{\odot}$ galaxies. These efficiencies are comparable to what was found in semi-analytic models by \citet{Choudhury:2008p2944} ($\epsilon = 0.008 - 0.009$ in the scenarios including dwarf galaxies) and \citet{Srbinovsky:2010p2800} ($\epsilon = 0.005$ in their fiducial model). 

Fig.~\ref{fig_reion} shows that a single high mass galaxy is a more efficient contributor to cosmic reionisation than a single low mass galaxy for the same set of parameters. This is consistent with the findings of \citet{Wise:2009p1755}, although they find slightly higher values for $10^{9} M_{\odot}$ galaxies, possibly due to their higher resolution. Even though the escape fractions we find are significantly lower than those found by \citet{Wise:2009p1755}, the stellar efficiency parameters are comparable. This is a result of the different phases in the lifetime of the galaxy that are targeted. \citet{Wise:2009p1755} studied primarily dwarf galaxies during the time of their assembly. During this phase the star formation rate and escape fraction are both high. However, because a large fraction of the gas in the galaxy has not been converted into stars yet, the stellar efficiency parameter is lowered. The phase we are looking at has lower escape fractions, but a larger fraction of the gas has been converted into stars, which raises the stellar efficiency parameter. We therefore find efficiency parameters that are consistent with those found by \citet{Wise:2009p1755}. 

Our findings suggest that high-redshift dwarf galaxies may be important for cosmic reionisation even after the initial starburst has died down and the galaxies have formed a rotationally supported disc. Although escape fractions are lower than in the starburst phase, the stellar fraction is much higher as a large fraction of the gas has been converted into stars. Therefore the efficiency parameter of the stars is only marginally lower than during the initial starburst. Our models suggest that high-redshift dwarf galaxies that are efficient in converting gas into stars, i.e., galaxies in a compact dark matter halo or with low spin parameter, or galaxies that have a low gas fraction but are still forming stars are potentially important contributors to cosmic reionisation. 


\section{Discussion}\label{section_discussion}

The escape fractions we find for high-redshift dwarf galaxies in this work differ considerably from those found in most previous work. \citet{Wise:2009p1755} found escape fractions ranging between 0.25 and 0.8 in AMR radiation hydrodynamics simulations of dwarf galaxies at redshift 8, using a ray-tracing algorithm to follow the radiation from the most massive sources. \citet{Razoumov:2010p1840} resimulated galaxies extracted from an SPH simulation of galaxy formation and found similar values for the escape fractions from low mass galaxies at high redshift, also with ray-tracing radiative transfer. For dwarf galaxies at $z \le 6$ that were extracted from a galaxy formation simulation \citet{Yajima:2011p2925} found escape fractions up to 70\%, although the authors report considerable scatter in the escape fraction from equal mass haloes. These values for the escape fraction are considerably higher than what we find in our models. On the other hand, using a moment radiative transfer method directly coupled to AMR hydrodynamics \citet{Gnedin:2008p2449} found in their simulations that it's impossible for radiation to escape from the dense inner parts of galaxies, resulting in almost no escape from low mass galaxies. In this section we discuss possible causes of the differences between our results and previous work.


\subsection{Resolution}

Contrary to previous work, in which galaxies were selected from cosmological simulation and thus had a given resolution, we are free to choose the resolution of our models. In the previous section we have shown that the resolution in our models is only sufficient to give a lower limit on the escape fraction. In principle this could explain the difference between our models and those of \citet{Wise:2009p1755}. However, the resolution in our models is higher than in the works by \citet{Razoumov:2010p1840} and \citet{Yajima:2011p2925}, while they report larger values for $f_{\mathrm{esc}}$ as well. Furthermore, the value for the escape fraction in our converged model is still significantly lower than what was found by \citet{Wise:2009p1755}. It therefore seems unlikely that the differences are solely caused by resolution effects.

The distribution of the ionising sources may play a role as well. We have shown in the previous section that low luminosity sources should not be neglected because their distribution in the galaxy is different from the more luminous, massive sources. This result is somewhat similar to the result of \citet{Gnedin:2008p2449}, who showed that the escape fraction from the disc galaxies in their simulation is dominated by old stars in the outer edges of the galaxy. We find that the low mass stars additionally provide channels for the radiation of the massive stars to escape. However, the works by \citet{Wise:2009p1755}, \citet{Razoumov:2010p1840} and \citet{Yajima:2011p2925} all find higher escape fractions than we do without including low mass sources. Is is therefore unlikely that the source distribution is responsible for this difference.


\subsection{The interplay of radiation and gas}

In this work we have post-processed the simulated galaxies with radiative transfer calculations, hence there is no direct coupling between the radiation and the gas in our models. \citet{Wise:2009p1755} showed that especially in low mass galaxies D-type ionisation fronts arising from radiative feedback are very effective in ejecting mass from a galaxy. However, the stellar feedback processes including UV-heating of the gas are incorporated in our model of the ISM and are therefore taken into account. Furthermore, \citet{Razoumov:2010p1840} found values for $f_{\mathrm{esc}}$ comparable to those found by \citet{Wise:2009p1755} without a direct coupling between radiation and gas. 

In all our runs the escape fraction converges within 5 Myr and usually even faster, around 2 to 3 Myr. This is well within the time that source evolution takes place, which is another indication that post-processing the hydrodynamics simulations does not affect the escape fraction calculations severely. We therefore think it is unlikely that the big difference in escape fractions is due to radiation hydrodynamics effects. We plan to investigate this issue further in future work.


\subsection{Physical processes}

The physical processes that are at work inside a galaxy play a crucial role for the escape fraction. For example, the star formation rate depends heavily on cooling processes and stellar feedback effects like UV heating, stellar winds and supernova explosions. The porosity of the ISM, that determines whether there are channels for the radiation to escape through, is highly dependent on supernova and stellar wind feedback. For the determination of the escape fraction it is therefore crucial to model the physics inside the galaxy in a realistic manner.

In contrast to previous work, the galaxy models in this study include the relevant physics to resolve the two-phase nature of the ISM in a self-consistent way. As star formation depends on the local properties of the ISM this is crucial for obtaining realistic star formation rates. It has been shown that our ISM model reproduces observed features of the gas distribution and star formation rates in local dwarf galaxies very well. We are therefore confident that the star formation rates presented in this work are realistic given the feedback parameters used. However, the choice for the IMF and the strength of the supernova feedback are important for the evolution of both the ISM and the galaxy as a whole. 

In all the models presented here we assume star formation to occur according to a Salpeter IMF. This is consistent with previous work, with the exception of the work by \citet{Wise:2009p1755}, who experiment with both a top-heavy and a Salpeter IMF. A top-heavy IMF results in the release of more ionising photons and an increase in the supernova rate. \citet{Wise:2009p1755} showed that a top-heavy IMF increases the escape fraction by 10-75\%. However, the escape fractions they find using a Salpeter IMF are still considerably higher than what we find in this work. As the works by \citet{Razoumov:2010p1840} and \citet{Yajima:2011p2925} also find high escape fractions using a Salpeter IMF, we conclude that the IMF is not the main cause of these differences.

In concordance with previous work we find that supernova feedback plays a key role in the escape of radiation by creating channels in the gas through which radiation can easily travel. We therefore expect that the escape fraction crucially depends on the strength of the supernova feedback and the supernova rate, the latter depending on the chosen IMF. In this work we assume stars above 8 $M_{\odot}$ to explode as type II supernova with an energy yield $\approx 10^{51}$ ergs. This is notably different from \citet{Wise:2009p1755} who keep the supernova feedback strength similar to that of a top-heavy IMF in all their runs. The different supernova feedback recipes may well be responsible for the large differences in escape fractions between our work and that of previous work. Given the importance of supernova feedback for the escape fraction, we plan to investigate the impact of different supernova recipes and IMFs in future work.   

In Sect.~\ref{section_timing} we have shown that the timing of the star formation and the local gas clearing plays an important role in the escape of radiation. If the peak of ionising luminosity occurs when stellar feedback has already influenced the surrounding gas, escape fractions are 50 to 100\% higher. This again shows that for numerical determinations of the escape fraction it is crucial to have realistic prescriptions the physical processes that shape the ISM around the sources. 


\subsection{Galaxy morphologies}

All the models we have studied are galaxies in isolation. In reality these galaxies form inside the filamentary structure of the cosmic web. Infalling gas from the filaments may lower the escape fraction by additional absorption, but may also have a positive effect by increasing the star formation rate. \citet{Wise:2009p1755} studied the difference between idealised haloes and cosmological haloes and the differences in escape fractions they found are too small to account for the differences in escape fractions between their and our work.

We have chosen to study the escape fraction from isolated galaxies because we focus on the physical processes that shape the galaxy and are of crucial importance for the amount of ionising radiation that is able to escape. The isolated initial conditions provide us with the means to impose the physical parameters of our choice on the galaxy, which enables us to gain insight in how the physical parameters of the galaxy influence the escape fraction. We concentrate on how the physical processes inside the galaxy shape the gas density and create the conditions for radiation to escape, assuming that outside influences have a less important effect. Given the strong dependence of the escape fraction on the shape of the ISM that we find this may not be a bad approximation. 

Another important difference between this work and previous work on high-redshift dwarf galaxies is that we study galaxies that have formed a rotationally supported disc, while \citet{Wise:2009p1755}, \citet{Razoumov:2010p1840} and \citet{Yajima:2011p2925} all studied the escape fraction from dwarf galaxies at the time of their assembly. When the galaxies assemble they go through an initial starburst phase, in which the infalling gas is converted into stars at very high rate. During this short initial phase one expects the star formation rate to be high and feedback effects to dominate, thus preventing the galaxy from forming a rotationally supported gas disc. Escape fractions are expected to be high in this evolutionary stage, which is indeed what these studies find. However, it is important to note that gas clearing effects are important in this stage as well, and it is unclear whether all previous studies have incorporated the physics to study this in necessary detail. 

In our models with low spin parameter, star formation rates are high and feedback effects shape the galaxies into highly irregular density distributions. The properties of these galaxies should therefore be comparable to the starburst phase studied in previous work. Indeed, the star formation in these models shows a more burst-like behaviour, and both the peak star formation rates and the escape fractions are close to the high values found in starbursting dwarf galaxies. 

After this starburst phase the galaxies form a rotationally supported disc. It is this phase that we capture in our simulations. In most of our model galaxies the evolution is much more quiet, which is reflected in the lower values of the escape fraction. Similar behaviour for disc galaxies, albeit with higher mass, was reported by \citet{Gnedin:2008p2449}. However, when the trend of decreasing escape fraction with galaxy mass from that work is extrapolated to the mass of the dwarf galaxies studied here, we find escape fractions that are higher. This is due to the fact that in the galaxies studied by \citet{Gnedin:2008p2449} almost no radiation escapes from within the galaxies, the escape fraction is dominated by stars in the outskirts. We find that due to our realistic treatment of the evolution of the ISM the gas inside the galaxies is shaped in such way that some of the radiation originating from inside the galaxy will be able to escape. It is important to note that the disc phase may be disrupted by infalling gas from filaments or mergers, leading to a significantly change in the gas distribution and star formation rate. We currently neglect these effects, but we plan to study this in future work.


\section{Conclusions}\label{section_conclusions}

We have studied the escape of ionising radiation from high-redshift dwarf galaxies with masses between $10^8$ and $10^{9} M_{\odot}$. By studying the galaxies in isolation we were able to determine the dependence of the escape fraction on the spin parameter, initial gas fraction and formation redshift. Contrary to previous work we have targeted the evolutionary stage in which the galaxies have formed a rotationally supported disc. We focussed on realistic treatment of the physical processes that shape the ISM to get accurate estimates of star formation in the galaxies and the column densities that ionising photons have to traverse in order to escape. Our results can be summarised as follows.
\begin{enumerate}
  \item We find mean escape fractions from high-redshift dwarf galaxies that lie between $10^{-7}$ and 0.01 for $10^{8} M_{\odot}$ galaxies and between $10^{-5}$ and 0.1 for $10^{9} M_{\odot}$ galaxies.
  \item Radiation preferentially escapes through holes blown by supernovae, therefore the escape is highly inhomogeneous.
  \item The escape fraction shows large variations over the lifetime of the galaxy, changing over 3 to 10 orders of magnitude.
  \item The escape of radiation is determined by the subtle interplay between opposing trends of lower escape fractions and higher star formation rates for higher gas fractions. We find no direct correlation between the mean star formation rate and the mean escape fraction.
  \item The angular momentum of the galaxy is important for the morphology of the galaxy. A low spin parameter results in a high escape fraction due to the high porosity of the ISM, which is in turn caused by more effective  supernova feedback.
  \item Dust is not important for the UV escape fraction. Including dust in the radiative transfer simulations changes the escape fraction by less than 1\%.
  \item The local gas complexes that give rise to star formation are the main constraint for the escape of radiation.
  \item For accurate determination of the escape fraction it is important to resolve the gas up to small scales to represent the porosity of the ISM correctly. Detailed modelling of the processes that shape the ISM and determine the star formation and feedback are essential for realistic numerical studies of the escape fraction.
  \item The distribution of the ionising sources in the galaxy plays an important role, excluding low luminosity sources and/or source merging will lead to incorrect calculations of the escape fraction.
  \item Although the absolute values of the escape fractions we find are low, our findings support the idea that high-redshift dwarf galaxies are important contributors to cosmic reionisation. We find that the efficiency parameter of the stars in our galaxy models is only slightly below the values needed in semi-analytic models of reionisation to reionise the Universe by redshift 6.
\end{enumerate}
The resolution in our models, while higher than most previous work, may not completely resolve the porosity of the ISM. The values of the escape fraction that we find should therefore be taken as a lower limit only. However, our resolution study shows that the converged value for the escape fraction is still below 1\%, thus our conclusion that radiation does not easily escape from the disc of a galaxy remains unchanged. 

External processes like gas accretion and merger events may have a profound impact on the number of ionising photons that are produced and escaping from a galaxy, for example by triggering a starburst. Thus, our assumption that the dwarf galaxies are isolated could lead to an underestimation of the escape fraction. On the other hand, dense filaments surrounding a galaxy may absorb a significant part of the ionising photons, resulting in a lower escape fraction. We plan to study this in more detail in future work.

In our models radiation primarily escapes through holes blown by supernovae in star forming regions. It may be possible to test this observationally by looking at the star formation surface density in the regions of the galaxy from where radiation escapes \citep{Siana:2010p2828}. The large variations of the escape fraction with time and angle suggest that observational studies need large samples to determine escape fractions. The inhomogeneous nature of the escape of radiation indicates that observations will find either very high escape fractions in galaxies for which radiation escapes in the line-of-sight or very low escape fractions for galaxies that have no channel for escaping radiation in the line-of-sight.

\begin{acknowledgements}
The authors would like to thank Garrelt Mellema and Eline Tolstoy for useful comments on the initial draft of this manuscript and the referee, John Wise, for comments that helped to clarify the content of this paper.
\end{acknowledgements}

\bibliographystyle{aa} 

\end{document}